\documentstyle[preprint,prb,aps,psfig,amssymb,amsbsy]{revtex}   % Uncomment this line for the preprint style
\begin{document}
%\preprint{}
\tightenlines
\draft
%\wideabs                                                         % Comment this line for the preprint style
{
\title{Internal Modes and Magnon Scattering on Topological Solitons\\ in 2d
Easy--Axis Ferromagnets} %
\author{Denis~D.~Sheka,\cite{DDSh_address} Boris~A.~Ivanov,$^\dag$ and
Franz~G.~Mertens} %
\address{Physikalisches Institut, Universit\"at Bayreuth,
D--95440 Bayreuth,  Germany\\
$^\dag$ Institute of Magnetism, NASU, 03142 Kiev, Ukraine}

\date{\today}

\maketitle

\begin{abstract}
We study the magnon modes in the presence of a topological
soliton in a 2d Heisenberg easy--axis ferromagnet. The problem of
magnon scattering on the soliton with arbitrary relation between
the soliton radius $R$ and the ``magnetic length'' $\Delta_0$ is
investigated for partial modes with different values of the
azimuthal quantum numbers $m$. Truly local modes are shown to be
present for all values of $m$, when the soliton radius is enough
large. The eigenfrequencies of such internal modes are calculated
analytically on limiting case of a large soliton radius and
numerically for arbitrary soliton radius. It is demonstrated that
the model of an isotropic magnet, which admits an exact analytical
investigation, is not adequate even for the limit of small radius
solitons, $R\ll\Delta_0$: there exists a local mode with nonzero
frequency. We use the data about local modes to derive the
effective equation of soliton motion; this equation has the usual
Newtonian form in contrast to the case of the easy--plane
ferromagnet. The effective mass of the soliton is found.
\end{abstract}
}

\section{Introduction} %
\label{sec:introduction}

Nonlinear topologically nontrivial excitations (solitons) are
well--known to play a special role in low--dimensional magnetic
systems. For example, the presence of vortices in 2d easy--plane
(EP) magnets gives rise to the
Berezinski\u\i--Kosterlitz--Thouless (BKT) phase
transition.\cite{Berezinsky72,Kosterlitz73} Kinks in 1d magnets
and localized Belavin--Polyakov (BP) solitons \cite{Belavin75} in
2d isotropic magnets are responsible for the destruction of
long--range order at finite temperature. The soliton signatures
in dynamical response functions can be observed experimentally.
Translational motion of solitons leads to the so--called soliton
central peak.\cite{Mikeska91,Wiesler94,Ivanov95e} Another
possibility to detect soliton signature is to look for magnon
modes, localized at the soliton ({\em local modes}, LM); such
modes are the cause of soliton magnetic resonance at the
characteristic frequencies of ``intrinsic''
motion.\cite{Boucher87}

This can be explained within the scope of the so--called soliton
phenomenology, where the magnet can be described as a
two--component gas of elementary excitations: solitons and
magnons. Such an approach was developed for 1d
magnets.\cite{Currie80} It was shown that the contribution of
magnons and soliton--magnon interaction is important for this
approach, as is obvious, for example, in the discussion of
soliton magnetic resonance.\cite{Currie80} But for 1d magnets the
soliton--magnon scattering causes a change of the magnon density
of states, which is necessary for a self--consistent calculation
of the temperature dependence of the soliton
density.\cite{Currie80}

For the 2d case, the concept of a soliton (vortex) gas has been
extended to describe EP magnets above the BKT vortex--antivortex
unbinding transition, with a finite density of
vortices.\cite{Mertens87,Mertens89} The concept of localized
topological solitons ($\pi_2$--topological solitons, see details
in Ref.\onlinecite{Kosevich90}) was used to explain the EPR
line--width in easy--axis (EA)
magnets.\cite{Waldner83,Waldner86,Waldner92,Zaspel93,Zaspel95,Zaspel96,Subbaraman98}
The classical example of a localized soliton is BP--solution,
which exists in isotropic magnets. The scattering problem for such
solitons allows an exact analytical solution because of the
scale-- and conformal-- invariance of the
model.\cite{Ivanov99,Walliser99} For EA magnets it is necessary
to take account of some important modifications due to the
breaking of such symmetries. That is why, both for  EA and EP 2d
magnets, the soliton density has not been calculated, but has
been used as an input parameter. The general features of the 2d
soliton dynamics are not clear at present. In particular, the
form of inertial terms in the dynamical equations for the soliton
centre is unknown; so the study of localized states is an issue
of current research. Note that the problem of the soliton
dynamics and the problem of the existence of LM  are intimately
connected to each other, and to the problem of soliton--magnon
scattering. For example, using numerical data for the scattering
amplitude, a non--Newtonian effective equation of motion of the
magnetic vortex was constructed.\cite{Ivanov98}

Usually non--1d solitons and, especially, the problem of
soliton--magnon scattering, are treated
numerically.\cite{Ivanov98,Wysin94,Wysin96} Only in the case of
purely isotropic magnets, with the exact BP--solution, the
analytical investigation of the scattering problem has been done
recently.\cite{Ivanov95g,Ivanov98a,Ivanov99} In particular, it
was found that a soliton has a number of local modes of zero
frequency;\cite{Ivanov95g} such modes determine the main features
of the scattering picture.\cite{Ivanov99} However, even a small
uniaxial anisotropy leads to principal changes. Truly local modes
(in contrast to quasi--local ones, below) with nonzero frequencies
were found numerically in Ref.~\onlinecite{Abdullaev99} for the
soliton in EA ferromagnets (FM) for some special values of the
azimuthal quantum number $m$; namely the mode with $m=-1$ for the
soliton with the topological charge $q=1$, and the mode with
$m=2$ for $q=2$ were investigated. Note that local modes do not
exist for the models of easy--plane and isotropic ferromagnets,
where the magnon spectrum has a gapless dispersion law. The case
of the 2d EP antiferromagnet is an exception; for this model a
finite--frequency, truly localized internal mode with $m=0$
exists inside the continuum spectrum.\cite{Ivanov96}

Recently there has also been renewed attention to the internal
dynamics of the soliton. It is caused by the studying of small
ferromagnetic particles in the so--called vortex state (magnetic
dots). Magnetic dots are micron--sized magnetic samples, placed
on a non--magnetic substrate; they have different shapes:
circular, quadratic, etc. The magnetic dots are interesting for
the usage in the high--density magnetic storage
devices\cite{Hillebrands97,Grimsditch98} and from a fundamental
point of view. As the dots are rather large (the size of magnetic
dot exceeds the critical size of a single domain), their ground
state could be inhomogeneous. Such a dot in the vortex state
contains a magnetic singularity in the centre of the vortex, or
the Belavin--Polyakov soliton type.\cite{Usov93} The discrete
magnon modes were observed for the uniformly magnetized dots by a
resonance technique;\cite{Mathieu98} their theory was constructed
using the Landau--Lifshitz equations\cite{Guslienko00}.

In this paper we consider the magnon modes which exist in 2d
Heisenberg EA FM with a topological soliton. The model of an
easy--axis ferromagnet is presented in Sec.~\ref{sec:model}. An
approximate solution for the soliton structure is examined. In
Sec.~\ref{sec:modes} we formulate equations, describing magnons
and solitons. In Sec.~\ref{sec:local} truly local magnon modes
are calculated analytically by variational approach for the
limiting cases of large and small radius solitons, and
numerically for arbitrary radii. In
Sec.~\ref{sec:scattering-numerical} the problem of
soliton--magnon scattering is formulated and treated numerically
for different partial modes. The phenomenology of a
pseudopotential is proposed in Sec.~\ref{sec:pseudopotential} to
analyze analytically both local states and the scattering
problem. In Sec.~\ref{sec:eq-mot} we use results for local modes
to derive the equations of soliton motion. Such equations are of
Newtonian type (with the account of gyroscopic force), in contrast
to both the case of the BP--soliton in the isotropic
magnet\cite{Ivanov99} and the case of the vortex in the EP
magnet\cite{Ivanov98}. In the framework of this equation the
soliton effective mass is calculated. A discussion and concluding
remarks are presented in Sec.~\ref{sec:conclusion}.

\section{The Model and Elementary Excitations} %
\label{sec:model}

The dynamics of the classical ferromagnet is described by the
Landau--Lifshitz equations for the normalized magnetization
$\boldsymbol m$. In angular variables, $ m_x+im_y=\sin \theta
\exp (i\phi )$, these equations correspond to the Lagrangian
\begin{equation} \label{eq:L}
L = A\int d^2x \left\{\frac 1D\left(1-\cos\theta\right)
\frac{\partial \phi}{\partial t}-{\cal W}(\theta,\phi)\right\}
\end{equation}
with the energy density $A\cdot {\cal W}(\theta,\phi)$,
\begin{equation} \label{eq:W}
 {\cal W}(\theta,\phi) =  \frac12\left( \nabla \theta
\right) ^2+\frac12\left( \nabla \phi \right) ^2\sin ^2\theta +
\frac1{2\Delta_0^2}\sin^2\theta,
\end{equation}
where $A=JS^2$, $J$ is the exchange integral, $S$ the atomic
spin, $\Delta_0=\sqrt{A/K}$ the characteristic scale (``magnetic
length''), $K$ the energy of the anisotropy, and $D$ the
spin--wave stiffness.

The model of EA FM has well-known magnon excitations
$\theta=\text{const}\ll1$, $\phi=\omega t-\boldsymbol{k}\cdot
\boldsymbol{r}$ above the homogeneous ground state. The dispersion
law for the magnons has a finite activation frequency
$\omega_0=D/\Delta_0^2$,
\begin{equation}\label{eq:dispersion_law}
\omega(\boldsymbol{k}) = \omega_0 + D|\boldsymbol{k}|^2,
\end{equation}
where $\boldsymbol{k}$ is the wave vector.

The simplest nonlinear excitation in the system is the dynamical
(precessional) topological soliton\cite{Kosevich90}
\begin{equation} \label{eq:dyn_soliton(1)}
\phi=\varphi _0 + q \chi + \Omega t,\; \theta=\theta_0(r),\;
\theta_0(0)=\pi, \theta_0(\infty)=0,
\end{equation}
where $r$ and $\chi$ are polar coordinates in the plane of the
magnet, the integer $q$ plays the role of a topological charge,
and $\Omega$ ($0<\Omega<\omega_0$) is the frequency of the
internal precession. The necessity to consider solitons with
internal precession is caused by the fact that, according to the
Derrick--Hobart theorem, static solitons in models like
(\ref{eq:W}) are unstable. Formally, there is no function
$\theta_0(r)$, which could provide a minimum of the energy
\begin{equation} \label{eq:energy}
E = \pi A\int_0^\infty rdr \left\{\left(
\frac{d\theta_0}{dr}\right) ^2 + \sin^2 \theta_0 \left(
\frac1{\Delta_0^2} + \frac{q^2}{r^2}\right) \right\}.
\end{equation}
The precessional soliton corresponds to the conditional minimum
of the energy (\ref{eq:energy}) for a given number $N$ of magnons
bound in the soliton
\begin{equation} \label{eq:N}
N = \frac{N_2}{\Delta_0^2}\int_0^\infty rdr
\left(1-\cos\theta_0(r)\right),
\end{equation}  which is conserved for the uniaxial case ($N_2=2\pi
A/\hbar\omega_0$ is the  characteristic number of bound magnons
in the 2d soliton),\cite{Voronov83} see also
Refs.~\onlinecite{Kosevich90,Bar'yakhtar93}. This corresponds to
the condition $\delta\left[E-N\hbar\Omega\right]=0$. The form of
the function $\theta_0(r)$ is defined by the ordinary
differential equation,
\begin{eqnarray}
&&\frac{d^2\theta_0}{dr^2} + \frac1r \frac{d\theta_0}{dr} -
\sin\theta_0 \cos\theta_0\left(\frac1{\Delta_0^2} +
\frac{q ^2}{r^2}\right)\nonumber\\
\label{eq:dyn_soliton(2)}
&&+\frac{\Omega}{\omega_0\Delta_0^2}\sin\theta_0 = 0,
\end{eqnarray}
with the boundary conditions (\ref{eq:dyn_soliton(1)}). The
precession frequency $\Omega$ is fully determined by the number of
bound magnons $N$. The dependence $\Omega(N)$ was calculated
numerically in Ref.~\onlinecite{Voronov83}; large radius solitons
($N\gg N_2$) have a very low precession frequency
($\Omega\ll\omega_0$); and $\Omega \to \omega_0$ for small radius
solitons  ($N\ll N_2$).

In the case of large radius solitons, $R\gg\Delta_0$, the
approximate ``domain wall'' solution\cite{Kosevich90}
\begin{equation}\label{eq:theta_r>>1}
\cos\theta_0(r) = \tanh \frac{r-R}{\Delta_0}
\end{equation}
is applicable, which describes a 1d domain wall at the distance
$R$ from the origin. One can obtain the relation between the
soliton radius $R$ and the frequency of the soliton precession
$\Omega$ from variational condition, in which  $R$ plays the role
of a fitting parameter for the  function (\ref{eq:theta_r>>1}).
In the first approximation $\Omega(R) \approx
\omega_0\cdot\Delta_0/R$.

The opposite case of small radius solitons, $R\ll\Delta_0$,
corresponds to the Belavin--Polyakov solution $\tan(\theta_0/2) =
(R/r)^{|q|}$, which describes the isotropic limit of the model
(\ref{eq:L}); it is valid for distances
$r\ll\Delta_0$.\cite{Belavin75} At larger distances the algebraic
decay is replaced by the general exponential form,
\begin{equation} \label{eq:theta_r<<1}
\theta_0(r)\propto\exp\left(-r/r_0\right),\qquad
r_0=\sqrt{\frac{D}{\omega_0-\Omega}}.
\end{equation}
The dependence $\Omega(R)$ was calculated both analytically and
numerically in Refs.~\onlinecite{Voronov83,Ivanov86}. The
frequency of the soliton precession satisfies $\Omega\to\omega_0$
when $R\to0$, but the dependence $\Omega(N)$ is not analytic, and
$d\Omega/dN\to\infty$ as $N/N_2\to0$.\cite{Voronov83}

\begin{figure}
\begin{center}
\psfig{width=3.5in,angle=0.0,file=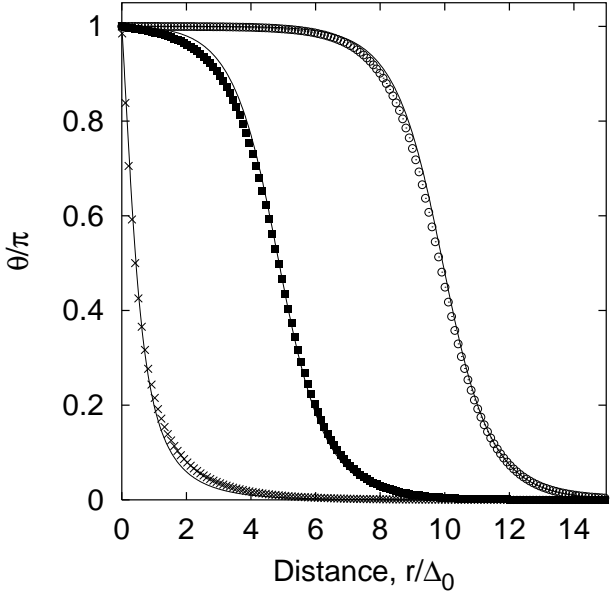}
\end{center}
\caption{Shape of $\theta_0(r)$ for different soliton radii
$R/\Delta_0$ ($\times=0.4$, $\blacksquare=5$, $\odot=10$). Lines:
numerical data, symbols: fitting data with the
trial function (\ref{eq:theta-trial})} %
\label{fig:structure}
\end{figure}

To describe approximately the soliton structure in the
intermediate case of arbitrary $R$, we use a fitting
method.\cite{Ivanov99b} We choose a trial function of the form
\begin{equation} \label{eq:theta-trial}
\tan\frac{\theta^{\text{(fit)}}_0(r)}{2} =
\frac{R^{\text{(fit)}}}{r}\cdot
\exp\left[-\frac{\left(r-R^{\text{(fit)}}\right)}{r_0}\right],
\end{equation}
which has valid asymptotics for $r\to0$ and $r\to\infty$ (we
discuss the case $q=1$ only). The parameter $R^{\text{(fit)}}$
can be found numerically by fitting the trial function
(\ref{eq:theta-trial}) to the numerical solution of
Eq.~(\ref{eq:dyn_soliton(2)}). Such an approximation shows that
the trial function describes the soliton shape with an accuracy of
about $10^{-2}$, see Table \ref{table:fitting}. From this table
one can see that the values of the fitting parameter
$R^{\text{(fit)}}$ are very close to the ``exact'' soliton radii;
it should be understood by the fact that
$\theta_0\left({r=R^{\text{(fit)}}}\right) = \pi/2$. Numerical
data for $\theta_0(r)$ and fitting results for
$\theta_0^{\text{(fit)}}(r)$ are plotted in
Fig.~\ref{fig:structure}.

\section{Magnon modes}
\label{sec:modes}

To analyze magnons on the soliton background, it is convenient to
introduce local coordinates $\left\{\boldsymbol{e}_1,
\boldsymbol{e}_2, \boldsymbol{e}_3 \right\}$, which describe the
configuration of the magnetization unit vector $\boldsymbol{m}$
in the unperturbed soliton: $\boldsymbol{e}_3$ coincides with
$\boldsymbol{m}$ in the soliton (\ref{eq:dyn_soliton(1)}),
$\boldsymbol{e}_1 =
\boldsymbol{e}_y\cos\phi_0-\boldsymbol{e}_x\sin\phi_0$, and
$\boldsymbol{e}_2 = \boldsymbol{e}_3 \times \boldsymbol{e}_1$.
Then the linear oscillations around the soliton can be described
in terms of projections of the magnetization $\boldsymbol{m}$ on
the local axes: $\mu=\boldsymbol{m}\cdot\boldsymbol{e}_1$ and
$\vartheta=-\boldsymbol{m}\cdot\boldsymbol{e}_2$. In the absence
of the soliton, $\theta_0=0$, $\phi_0=\Omega t$, such oscillations
correspond to free  magnons in the rotating coordinate frame.

The linearized equations for $\vartheta$ and $\mu$ can be
presented in the form:
\begin{mathletters}\label{eq:vartheta&mu}
\begin{eqnarray}\label{eq:vartheta}
&&\left[ - \nabla^2 + {\cal U}_1(r) \right] \vartheta + \frac{2q
\cos\theta_0}{r^2}\frac{\partial \mu}{\partial \chi} =
\frac1{\omega_0\Delta_0^2} \frac{\partial\mu}{\partial t},
\\
\label{eq:mu} %
&&\left[ - \nabla^2 + {\cal U}_2(r) \right] \mu - \frac{2q
\cos\theta_0}{r^2}\frac{\partial \vartheta}{\partial \chi} = -
\frac1{\omega_0\Delta_0^2}\frac{\partial\vartheta}{\partial t}
\end{eqnarray}
\end{mathletters}
with the ``potentials''
\begin{mathletters}\label{eq:U1&U2}
\begin{eqnarray}\label{eq:U1}
{\cal U}_1(r) &=& \left(\frac1{\Delta_0^2} + \frac{q
^2}{r^2}\right)\cos2\theta_0(r) -
\frac\Omega{\omega_0\Delta_0^2}\cos \theta_0(r),
\\
\label{eq:U2} %
{\cal U}_2(r) &=& - \left(\frac{d\theta_0(r)}{dr}\right)^2 +
\cot\theta_0(r)\nabla_r^2\theta_0 \nonumber\\ &=& -
\left(\frac{d\theta_0(r)}{dr}\right)^2 + \left(
\frac1{\Delta_0^2} + \frac{q
^2}{r^2}\right)\cos^2\theta_0(r)\nonumber\\ && -
\frac\Omega{\omega_0\Delta_0^2}\cos \theta_0(r),
\end{eqnarray}
\end{mathletters}
where $\nabla_r^2$ is the radial part of the Laplace operator.

\begin{figure}
\begin{center}
\psfig{width=3.5in,angle=0.0,file=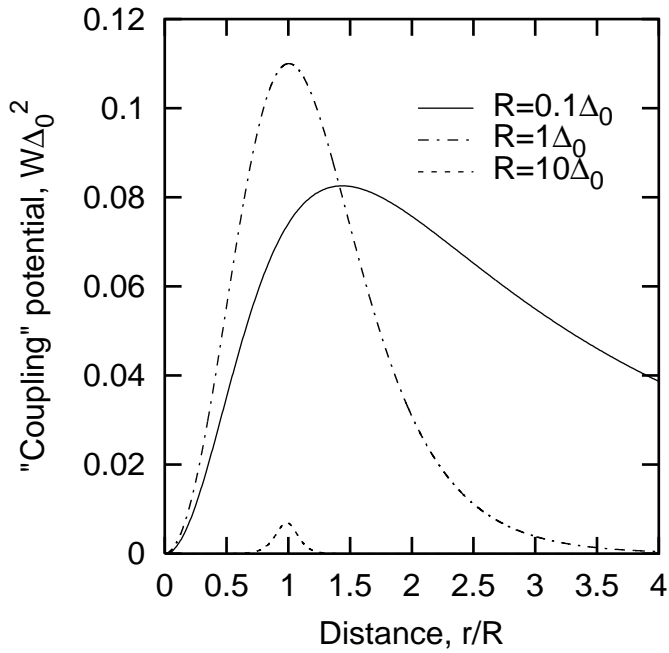}
\end{center}
\caption{Shape of the ``coupling'' potential for different soliton
radii} \label{fig:W}
\end{figure}

It is convenient to represent the solutions of
Eqs.~(\ref{eq:vartheta&mu}) in terms of a partial--wave expansion
for $\vartheta$ and $\mu$
\begin{mathletters} \label{eq:vartheta&mu-via-u&v}
\begin{eqnarray} \label{eq:vartheta-via-u&v}
\vartheta &=&\sum_{\alpha} \Bigl( u_\alpha(r) + v_\alpha(r)
\Bigr)\cos(m\chi+\tilde{\omega}_\alpha t)\\
\label{eq:mu-via-u&v} %
\mu &=&\sum_{\alpha} \Bigl( u_\alpha(r) - v_\alpha(r)
\Bigr)\sin(m\chi+\tilde{\omega}_\alpha t),
\end{eqnarray}
\end{mathletters}
where $\alpha=(k,m)$ is a full set of eigennumbers, $k$ and $m$
being radial and azimuthal quantum numbers, respectively. The
quantity $\tilde{\omega} = \omega-\Omega$ is the magnon frequency
in the rotating frame; $k$ and $\tilde{\omega}$ are connected by
the dispersion law, cf. Eq.~(\ref{eq:dispersion_law}),
\begin{equation}\eqnum{\ref{eq:dispersion_law}$'$}\label{eq:dispersion_law'}
\tilde{\omega}(\boldsymbol{k}) = \omega_{\text{gap}} +
D|\boldsymbol{k}|^2, \qquad \omega_{\text{gap}} = \omega_0 -
\Omega,
\end{equation}
where $\omega_{\text{gap}}$ is the magnon gap frequency in the
rotating frame. Below we will replace $\tilde{\omega}$ by
$\omega$, almost everywhere without confusion, because we will
work in the rotating frame mainly; the index $\alpha$ will be
omitted, too. Thus we then finally obtain the following eigenvalue
problem (EVP) for the functions $u$ and $v$:
\begin{mathletters} \label{eq:u&v}
\begin{eqnarray} \label{eq:u&v(1)}
\hat{H}_1 u & \equiv & \left[ \hat{H}_0 +\frac{m^2}{r^2} + V(r)
- \frac{{\omega}}{\omega_0\Delta_0^2} \right] u = W(r) v,\\
\label{eq:u&v(2)} %
\hat{H}_2 u & \equiv &  \left[ \hat{H}_0 +\frac{m^2}{r^2} - V(r)
+ \frac{{\omega}}{\omega_0\Delta_0^2} \right] v = W(r) u,
\end{eqnarray}
\end{mathletters}
where $\hat{H}_0=-\nabla_r^2+U_0(r)$ is the 2d radial
Schr\"{o}dinger operator; the ``potentials'' are
\begin{mathletters} \label{eq:potentials}
\begin{equation}\label{eq:U0,V}
U_0(r)=\frac12\left({\cal U}_1 + {\cal U}_2\right), \qquad V(r)
=\frac{2qm\cos\theta_0}{r^2}.
\end{equation}
Note, that the Eqs.~(\ref{eq:u&v}) are invariant under the
conjugations ${\omega}\to-{\omega}$, $m\to-m$, and
$u\leftrightarrow v$. In a classical theory we can choose any
sign of the frequency; but in order to make contact with quantum
theory, with a positive frequency and energy ${\cal E}_k=\hbar
|\omega_k|$, we will discuss the case ${\omega}>0$ only. Thus,
there are two different sets of Eqs.~(\ref{eq:u&v}) for $m=|n|$
and $m=-|n|$. Besides there appears a difference between
Eqs.~(\ref{eq:u&v(1)}) and (\ref{eq:u&v(2)}). Since
(\ref{eq:u&v(2)}) has the asymptotically equivalent form,
$-\nabla_r^2v+\omega v=0$, the function $v$ has an exponential
behaviour; at the same time Eq.~(\ref{eq:u&v(1)}) yields
oscillating solutions. So, by choosing $\omega>0$, the variable
$v$ becomes a slave variable in the EVP (\ref{eq:u&v}).

The ``coupling potential'' $W$,
\begin{eqnarray}
W(r) &=& \frac12 \left({\cal U}_2 - {\cal U}_1 \right) \nonumber \\
\label{eq:W(r)} &=& \frac12\left[\sin^2\theta_0 \left(
\frac1{\Delta_0^2} +\frac1{r^2}\right) -
\left(\frac{d\theta_0}{dr}\right)^2 \right],
\end{eqnarray}
\end{mathletters}
is positive, it has a maximum value at a distance of about the
soliton radius $R$ from the origin, vanishing both at $r=0$ and
far from the soliton. Using the ``domain wall'' approximation
(\ref{eq:theta_r>>1}), one obtains that the maximum value of
$W\propto 1/R^2$, when $R\gg\Delta_0$. In the opposite case of
small $R$, in the range $r\leq R$ the soliton looks like the
Belavin--Polyakov solution; here $r(d\theta_0/dr)\approx -
\sin\theta_0$, and $W=0$ in the main exchange
approximation.\cite{Ivanov99} Therefore, one can suppose that
$W\ll 1/\Delta_0^2$ for any $R$. In fact, this is confirmed by
numerical calculations, see Fig.~\ref{fig:W}.

However, one can see below in Sec.~\ref{sec:local} that the
problem is impoverished when the ``coupling potential'' $W(r)$
is  ignored; even in the limits of small (large) soliton radii,
the main properties of local modes vanish, if we assume that
$W(r)=0$. The reason is that a set of more than one coupled
Schr\"{o}dinger--like equations has properties, which are absent
from a single equation. To explain this fact, remember that for a
single 1d Schr\"{o}dinger equation there exist two independent
solutions; for the eigenfrequencies inside the continuous
spectrum such solutions have oscillating asymptotics. For a set
of 2, 3, \ldots\ equations there exist 4, 6, \ldots\ linearly
independent solutions, respectively; some of them can have
exponential asymptotics. Naturally, the real modes have an
oscillational form here; we will use this fact below for the
numerical analysis. But it is necessary to take into account the
exponential solutions, too. In particular, as was shown in
Ref.~\onlinecite{Ivanov96}, such exponential asymptotics could
correspond to truly localized states inside the continuous
spectrum, which are forbidden for equations of the Schr\"{o}dinger
form. In the reduced EVP with two independent equations like
(\ref{eq:u&v}) with $W=0$, there is a continuous spectrum
${\omega}>\omega_{\text{gap}}$ for the function $u(r)$ ($u\neq0$,
$v=0$) and ${\omega}<-\omega_{ \text{gap}}$ for the function
$v(r)$ ($u=0$, $v\neq0$). There are no limitations for the lowest
value of the discrete levels; in particular, the frequencies of
the discrete spectrum could be less then $-\omega_{\text{gap}}$
for the function $u$. At the same time, as follows from the
general set of coupled Eqs.~(\ref{eq:u&v}), frequencies of local
modes lie in the region $-\omega_{\text{gap}}<{\omega}<
\omega_{\text{gap}}$. Therefore, it is necessary to solve the
general EVP (\ref{eq:u&v}) for an exact analysis of the problem.

We should note here that the model of an isotropic magnet is a
special case, in which the ``coupling'' potential is exactly
equal to zero. To study an EVP like (\ref{eq:u&v}) with $W=0$, it
is sufficient to take into account one equation only (namely, the
equation for the variable $u$); the second one, which has
unphysical solutions for $\omega>0$, was ignored in
Ref.\onlinecite{Ivanov99}. That is why the results for EA FM,
obtained in this article, and the results for the isotropic
magnet\cite{Ivanov99} differ strongly even for small anisotropy.
One difference is  caused by the presence of truly local modes.
Another one is due to the peculiar structure of the
quasi--continuous spectrum with ${\omega}> \omega_{\text{gap}}$
for a finite geometry: the spectrum of the EA FM has a system of
doublets with opposite signs of $m$, while there are no doublets
for the isotropic magnet.

\section{Local modes}
\label{sec:local}

According to the dispersion law (\ref{eq:dispersion_law'}), free
magnons only exist when ${\omega}>\omega_{\text{gap}}$. It is
naturally to look for local modes in the range
$0<{\omega}<\omega_{\text{gap}}$ (remember that we discuss the
case ${\omega}>0$ ). In principle, such local modes could exist
inside the continuous spectrum, as for the case of the nonlinear
$\sigma$--model, describing the EP antiferromagnet.\cite{Ivanov96}
However, our study shows that this is not the case for the EA FM.
Existence of local modes in such a magnet was found numerically in
Ref.~\onlinecite{Abdullaev99} for a soliton with some special
parameters; namely the mode with $m=-1$ for the soliton with the
topological charge $q=1$, and the mode with $m=2$ for $q=2$ were
calculated. Such modes appear in field theories, which include
soliton--like solutions.\cite{Walliser99}

To study local modes, note that near the soliton centre,
$r\ll\Delta_0$, one can obtain the asymptotics of the magnon
amplitudes in the following form:
\begin{mathletters}\label{eq:u&v4x<<1}
\begin{eqnarray}\label{eq:u4x<<1}
u_m(r) &=& {\cal A}_m \left(\frac{r}{\Delta_0}\right)^{|q-m|}
\left[1+O\left({r^2}/{\Delta_0^2}\right)\right], \\
\label{eq:v4x<<1} v_m(r) &=& \varepsilon {\cal A}_m
\left(\frac{r}{\Delta_0}\right)^{|q+m|}
\left[1+O\left({r^2}/{\Delta_0^2}\right)\right].
\end{eqnarray}
\end{mathletters}
Here the coefficient ${\cal A}$ can be determined by the
normalization; the presence of a nonzero factor $\varepsilon$ is
caused by the ``interaction'' $W(r)$ between
Eqs.~(\ref{eq:u&v(1)}) and (\ref{eq:u&v(2)}); its value cannot be
found through this  asymptotic expansion.

In the opposite case of large distance, $r\gg\Delta_0$, the EVP
(\ref{eq:u&v}) has the following general solution (here and below
we consider the most interesting case of solitons with unit
topological charge $q=1$, which has lowest energy)
\begin{mathletters} \label{eq:u&v-loc-infty}
\begin{eqnarray} \label{eq:u-loc-infty}
u_m(r)&\approx& A_mK_{|m+1|}(\varkappa_1^-r) + B_m I_{|m+1|}(\varkappa_1^-r)\\ %
\label{eq:v-loc-infty} %
v_m(r)&\approx& C_mK_{|m-1|}(\varkappa_1^+r) + D_m I_{|m-1|}(\varkappa_1^+r),%
\end{eqnarray}
\end{mathletters}
where $K_\nu$ and $I_\nu$ are Macdonald and modified Bessel
functions respectively, and $\varkappa_1^\pm =
\sqrt{\left(\omega_0-\Omega \pm{\omega}\right)/ D}$. Note, that
at large distances, $z\gg1$, the modified Bessel functions have
exponential behaviour, $I_\nu(z) \propto \exp(z)/\sqrt{z}$,
$K_\nu(z) \propto \exp(-z)/\sqrt{z}$, so we use the asymptotical
condition $u_m, v_m \propto \exp(-\varkappa_1^\pm r)/\sqrt{r}$ to
obtain the localized solutions.

Firstly note that for the special cases $m=0,+1$ there exist
``zero--modes'' with ${\omega}=0$. Such modes are caused by the
internal symmetry of the problem. They have the same form as for
EP magnets.\cite{Ivanov98} One of the zero--modes, the so--called
translational mode with $m=+1$,
\begin{mathletters} \label{eq:m=-1,0,+1}
\begin{equation}\label{eq:m=+/-1}
u_{+1}=\frac{d\theta_0}{dr} - \frac{\sin\theta_0}{r}, \qquad
v_{+1}=\frac{d\theta_0}{dr} + \frac{\sin\theta_0}{r}
\end{equation}
describes the position shift of the soliton. Another one with
$m=0$
\begin{equation}\label{eq:m=0}
u_0 = \sin\theta_0,\qquad v_0 = -\sin\theta_0
\end{equation}
\end{mathletters}
is a rotational mode, which corresponds to the presence of an
arbitrary parameter, the phase $\varphi_0$, in the soliton
structure (\ref{eq:dyn_soliton(1)}). For the isotropic magnet, as
well as for the EP magnet, similar modes exist, too; but there
they are quasi--local due to their slow power--law decay, $1/r$,
or $1/r^2$. In the EA magnets the decay has an exponential form.

\begin{figure}
\psfig{width=3.5in,angle=0.0,file=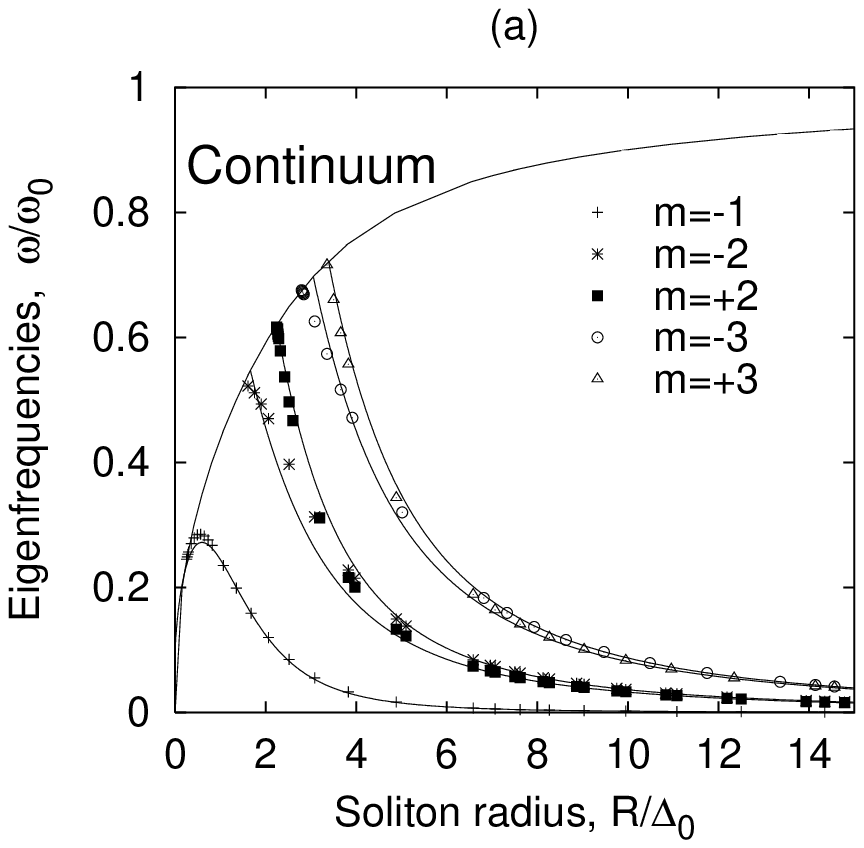}
\psfig{width=3.5in,angle=0.0,file=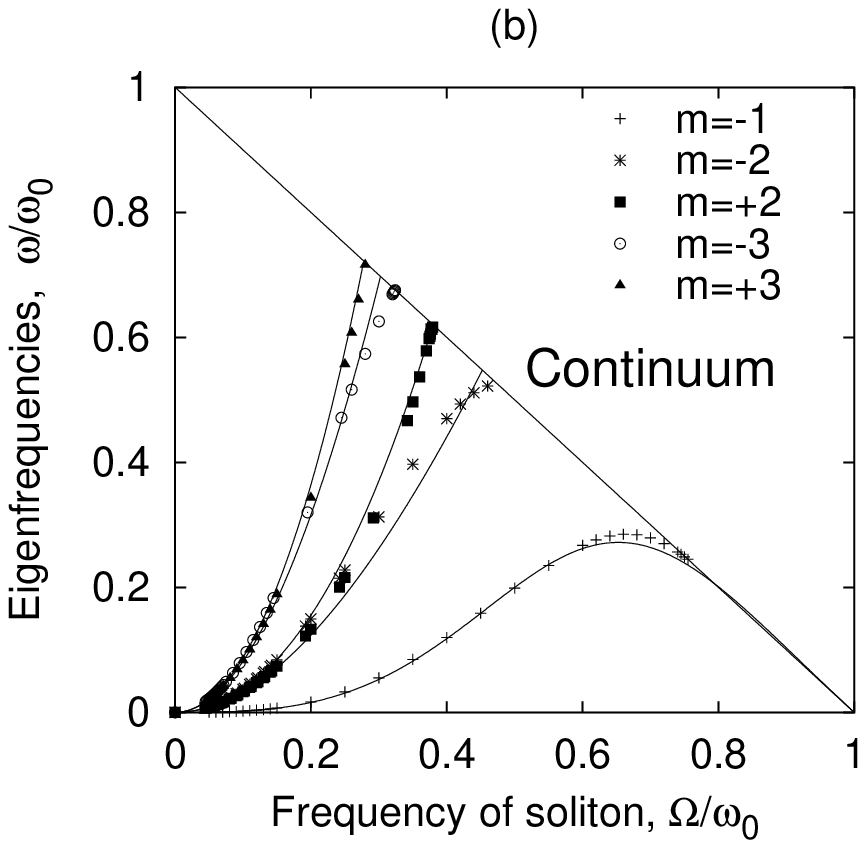}
\caption{Eigenfrequencies of local modes as functions of: (a)
soliton radius and (b) frequency of internal precession of the
soliton. Lines: theoretical results from
Eqs.~(\ref{eq:omega_m(R>>1)}) and (\ref{eq:omega:m=-1(trial)});
symbols: numerical data.} \label{fig:omega}
\end{figure}

The remaining modes could be investigated numerically; analytical
results were obtained for some limiting cases only, see below. To
study the EVP numerically, we use the two--parameter shooting
method for the integration of the set of Eqs.~(\ref{eq:u&v}); a
similar one--parameter shooting scheme was proposed in
Ref.~\onlinecite{Ivanov96} to study the vortex--magnon scattering
problem in 2d antiferromagnets. Numerically, we ``kill'' two
growing exponents in (\ref{eq:u&v-loc-infty}) by choosing two
shooting parameters $\varepsilon$ and $\omega$.  As a result, we
have found numerically that there are a number of local modes.
Modes with opposite signs of $m$ bind to doublets; this picture is
well--pronounced for $|m|>1$ in the limit of large $R$, see
Fig.~\ref{fig:omega}.

Such local modes exist only for not too small soliton radii (not
for large $\Omega$). Eigenfrequencies for such modes with
arbitrary azimuthal numbers $m$ ($|m|>1$) increase rapidly with
the decreasing of the soliton radius, and reach the boundary of
the continuous spectrum. It looks as if all such modes ``leave''
the region of the gap in the spectrum, if the soliton radius is
smaller than some critical value $R_m^{\text{c}}$, see Table
\ref{table:Rc}. It is natural to suppose that all modes with
$|m|>1$ transform to the quasi--local ones with decreasing of
$R$; this fact can be tested numerically due to the singularities
of the scattering amplitudes on such quasi--local modes, see the
next section.

Thus, in the range $R \lesssim 1.52\Delta_0$ there is only one
local mode with non--zero frequency, $m=-1$, see
Fig.~\ref{fig:omega}. That is why only the mode with $m=-1$ was
observed in Ref.~\onlinecite{Abdullaev99} for the soliton with
$q=1$.

\begin{figure}
\psfig{width=3.5in,angle=0.0,file=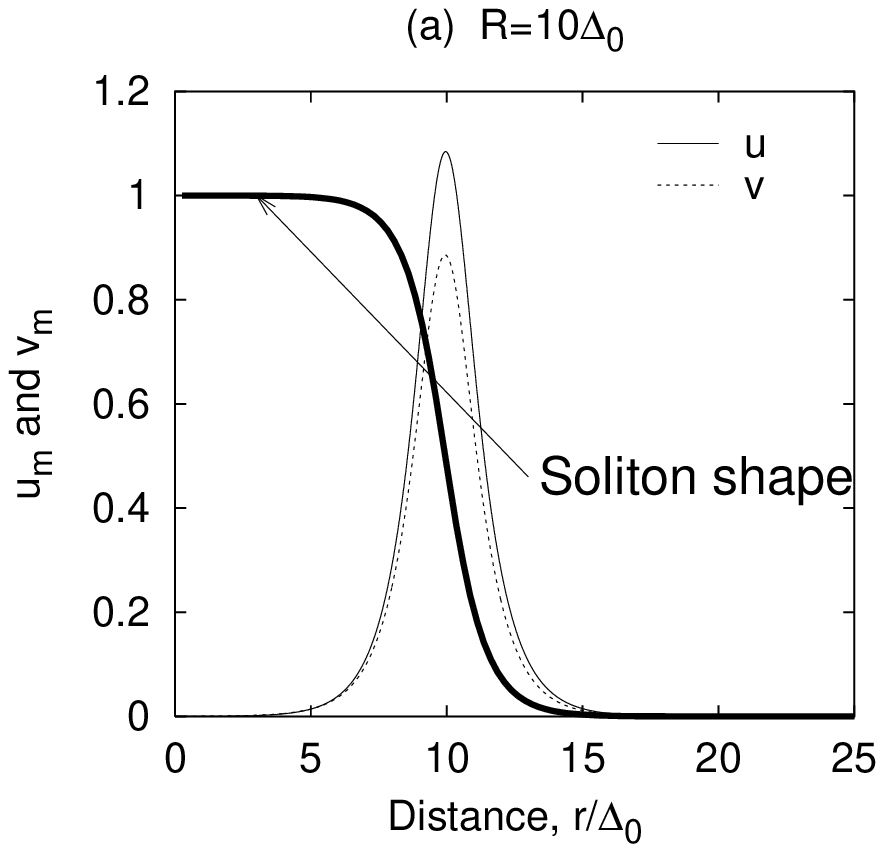}
\psfig{width=3.5in,angle=0.0,file=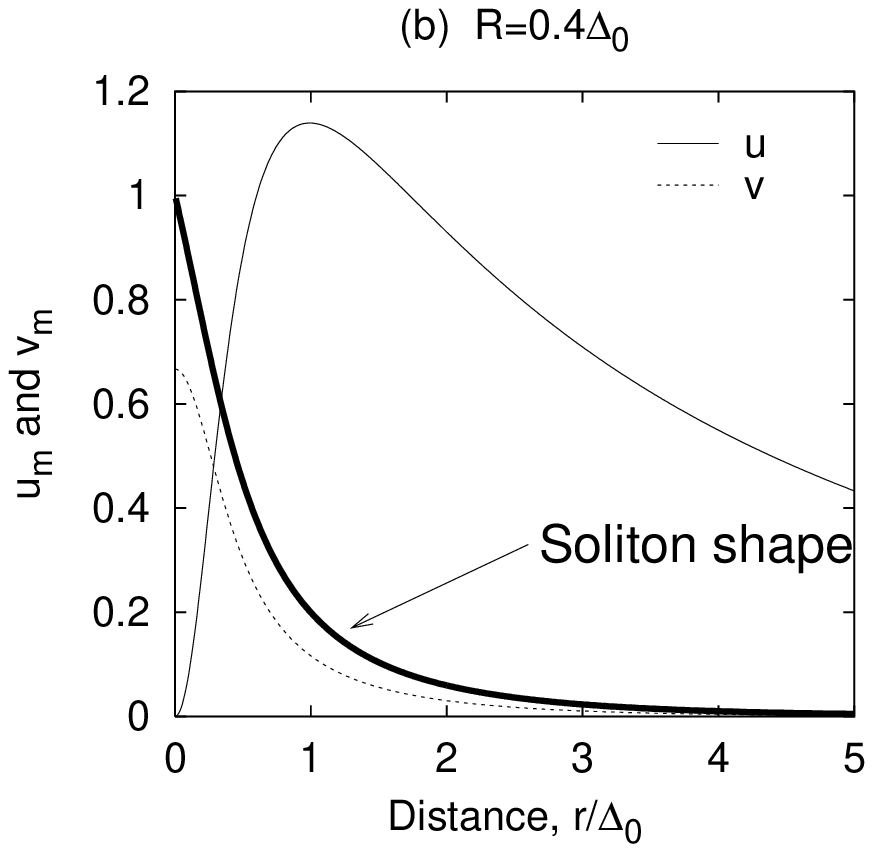}
\caption{Magnon amplitudes for local modes with $m=-1$}
\label{fig:local_R}
\end{figure}

The eigenfunctions of bound states $u$ and $v$ are localized near
the soliton core. The functions $u$ and $v$ are plotted in
Fig.~\ref{fig:local_R} for the mode with $m=-1$; the picture of
well--localized eigenfunctions is specially pronounced for the
large radius solitons, see Fig.~\ref{fig:local_R}(a).

In order to analyze the spectrum of local modes analytically, let
us reformulate the EVP (\ref{eq:u&v}) as a variational problem
for the functional
\begin{equation} \label{eq:functional}
F[u,v] = \int_0^\infty rdr\left(u \hat{H}_1 u + v \hat{H}_2 v -
2Wuv \right).
\end{equation}

We start with large radius solitons, $R\gg\Delta_0$, when the
``domain wall'' approximation (\ref{eq:theta_r>>1}) is valid. In
the region of interest, $r\simeq R\gg\Delta_0$, the operator
$\hat{H}_0$ from the EVP (\ref{eq:u&v}) has in the main
approximation for $\Delta_0/R$ a simple reflectionless potential,
$$\hat{H}_0 \approx \hat{L}\equiv-\frac{d^2}{dr^2} +
\frac1{\Delta_0^2}\left[1-\frac2{\cosh^2\left((r-
R)/\Delta_0\right)}\right]$$ with a single localized state
$\psi_0$, corresponding to the zeroth eigenvalue
$$\hat{L}\psi_0=0,\quad \psi_0(r)= \left(2R\Delta_0\right)^{-1/2}/
\cosh\left((r-R)/\Delta_0\right).$$ The other potentials in the
EVP can be considered as corrections to $\hat{L}$; this is the
reason to put the trial functions in the form $u = a\psi_0$, $v =
b\psi_0$ with trial parameters $a$ and $b$. The variational
approach with such simple trial functions leads to
eigenfrequencies for the different modes in the form
\begin{eqnarray} \label{eq:omega_m}
&&{\omega}_m = m\alpha+\sqrt{\left(m^2\beta-\delta\right)^2-\gamma^2},\\
&& \alpha = \left\langle \psi_0\left|
\frac{2\cos\theta_0(r)}{r^2} \right| \psi_0\right \rangle,\qquad
\beta=\left\langle \psi_0\left| \frac{1}{r^2}
\right| \psi_0\right\rangle,\nonumber \\
&& \gamma = \left\langle \psi_0\left| W(r) \right| \psi_0
\right\rangle, \; \delta = \left\langle \psi_0\left|
\frac1r\frac{d}{dr} + \Omega\cos\theta_0(r)\right| \psi_0
\right\rangle\nonumber
\end{eqnarray}
where $\langle \psi_0\left| \dots \right| \psi_0 \rangle =
\int_0^\infty \psi_0 (\dots) \psi_0 rdr$. However, there is a
difference between the real solutions $u$, $v$, and the trial
function $\psi_0$. This difference provides, e.g., the existence
of ``zero--modes''. We do not like to improve the form of the
eigenfunctions, but only the coefficients in the
Eqs.~(\ref{eq:omega_m}), which describe the eigenfrequencies.
Using the additional condition for ``zero--modes'',
\begin{equation} \label{eq:omega=0}
{\omega}_m=0 \qquad \text{for $m=0,+1$},
\end{equation}
the eigenfrequencies $\omega_m$ can be expressed through two
independent coefficients only. It is convenient to present the
eigenfrequencies in the form
\begin{equation}\label{eq:omega_m*}
{\omega}_m = m\alpha+|m|\sqrt{\alpha^2+\beta^2(m^2-1)}.
\end{equation}
Using trial functions $\psi_0$, we can calculate the parameters
$\alpha$ and $\beta$ for the large radius solitons,
\begin{mathletters} \label{eq:omega_m_cases}
\begin{equation} \label{eq:alpha&beta}
\alpha=-\omega_0(\Delta_0/R)^3 \qquad
\beta=\omega_0(\Delta_0/R)^2.
\end{equation}
In this case the spectrum has the form
\begin{equation}\label{eq:omega_m(R>>1)}
\frac{{\omega}_m}{\omega_0} =
|m|\sqrt{\left(\frac{\Delta_0}{R}\right)^{\!6} +
\frac{m^2-1}{(R/\Delta_0)^4}}-
m\left(\frac{\Delta_0}{R}\right)^{\!3}
\end{equation}
This formula corresponds very well to the numerical data for
$R\gg\Delta_0$, see Fig.~\ref{fig:omega}. In the main
approximation for $\Delta_0/R$ there is a system of doublets (for
all modes with $|m|>1$) with quadratic dependence of the mean
frequency on $\Delta_0/R$,
\begin{equation}\label{eq:omega_doublets}
\overline{{\omega}}_{\pm m}\approx\omega_0|m|\sqrt{m^2-1}
\left(\frac{\Delta_0}{R}\right)^2,
\end{equation}
and the splitting is a small cubic correction, $\Delta\omega_{\pm
m}\propto m\left(\Delta_0/R\right)^3$. For the special case $m=-1$
the dependence $\omega(R)$ has the form
\begin{equation}\label{eq:omega_m=-1}
{\omega}_{m=-1} \approx
2\omega_0\left(\frac{\Delta_0}{R}\right)^3.
\end{equation}
\end{mathletters}

Eq.~(\ref{eq:omega_m(R>>1)}) describes qualitatively all modes
with arbitrary $R$. In particular, it describes the fact,
mentioned above, that modes  reach the boundary of the continuous
spectrum at some finite frequency of the soliton precession
$\Omega_m^{\text{c}}$. Let us estimate the positions of the
crossover points, where the modes ``leave'' the discrete spectrum.
Using Eq.~(\ref{eq:omega_m(R>>1)}), one can obtain the condition
for the critical values of $x=\Omega_m^{\text{c}}/\omega_0$:
\begin{equation} \label{eq:Omega^c-est}
|m|\sqrt{x^6+(m^2-1)x^4}-mx^3=1-x.
\end{equation}
Results are very close to the numerical data, see Table
\ref{table:Rc}. We expect that the local modes transform into the
quasi--local ones inside the continuous spectrum; but their
frequencies can still be described by the formula
(\ref{eq:omega_m(R>>1)}), see Sec.~\ref{sec:pseudopotential}.

For the small radius solitons, $R\ll\Delta_0$, there only exist
local modes with $|m|=1$. Thus, we will consider this case.
Constructing the trial functions, one can suppose that they
should be similar to the modes with $m=\pm1$. Eigenfunctions for
the translational ``zero--mode'', $u_{+1}$ and $v_{+1}$, are
well--known, see Eq.~(\ref{eq:m=+/-1}). To construct the trial
functions it is convenient to use functions $u_{-1}$ and $v_{-1}$
in the form
$$u_{-1}=r^2u_{+1},\qquad v_{-1}=r^2v_{+1},$$ which coincide with the
exact ``zero--mode'' eigenfunctions for the mode with $m=-1$ in
the limiting case of the BP--soliton.\cite{Ivanov99} Now we can
choose the trial functions $u$ and $v$ as follows,
\begin{equation} \label{eq:u+v4smallR}
u = a\cdot u_{+1}+b\cdot u_{-1}, \quad v = c\cdot v_{+1}+d\cdot
v_{-1.}
\end{equation}
The variational problem (\ref{eq:functional}) can be solved
analytically in the main approximation for $R/\Delta_0$ using the
explicit BP--solutions. Tedious calculations lead to the
eigenfrequency for small $R$,
\begin{equation} \label{eq:omega_m=-1_small_R}
{\omega}_{m=-1} \approx C\cdot\left(\omega_0-\Omega\right).
\end{equation}
Unfortunately, this method is not exact and leads to a value of
the constant $C$ which is larger than the numerical value. This
is caused by the fact that the trial functions, which are
constructed from zero modes, have the wrong asymptotical behaviour
(\ref{eq:theta_r<<1}) at large distance, $\theta_0(r)$,
$\theta'_0(r)$ $\propto\exp(-r/r_0)$, instead of
(\ref{eq:u&v-loc-infty}), i.e. $u \propto \exp(-\varkappa_1^-r)$,
$v \propto \exp(-\varkappa_1^+r)$. To improve this result, one
can change the asymptotical behaviour of the functions
(\ref{eq:u+v4smallR}) by exponential factors:
$$ u \longrightarrow u\cdot e^{r\left(1/r_0-\varkappa_1^-
\right)},\qquad v \longrightarrow v\cdot
e^{r\left(1/r_0-\varkappa_1^+\right)}.$$ In this case the
calculations can be done numerically only; the value of the
constant $C\approx1$. For the approximate description of the
eigenfrequency of the mode with $m=-1$  in the intermediate case
of arbitrary $R$, we use the trial function
\begin{equation} \label{eq:omega:m=-1(trial)}
\omega_{m=-1}^{(trial)}(\zeta)=\omega_0\cdot
\frac{2\zeta^3(1-\zeta)}{1-\zeta+2\zeta^4}, \qquad
\zeta=\frac{\Omega}{\omega_0}.
\end{equation}
Providing the asymptotically correct behaviour,
(\ref{eq:omega_m=-1}) and (\ref{eq:omega_m=-1_small_R}), this
approximation reproduces to the numerical results with an
accuracy of about $2\cdot10^{-3}$, see Fig.~\ref{fig:omega}.

\section{The scattering problem: numerical results}
\label{sec:scattering-numerical}

Let us describe the scattering of magnons by a soliton. Note that
without a soliton, free magnon modes have the form
\begin{equation} \label{eq:modes_free}
u_m(r) \propto J_m(kr), \qquad v_m(r) = 0,
\end{equation}
where $k$ is a ``radial wave number'', $k$ and ${\omega}$ are
connected by the dispersion law (\ref{eq:dispersion_law'}), and
$J_m$ are Bessel functions. The free--modes like $u_m$ play the
role of the partial cylinder waves of a plane spin wave
\begin{equation} \label{eq:plain_wave*}
\exp\left(i\boldsymbol{k}\cdot\boldsymbol{r}-i\omega t\right) =
\sum_{m=-\infty}^\infty i^mJ_m(kr)e^{im\chi-i\omega t}.
\end{equation}

To describe magnon solutions in the presence of a soliton, one
should note that far from the soliton, $r\gg\Delta_0$, the
``coupling potential'' $W(r)$ is exponentially small. Therefore
there are two asymptotically independent EVP; the magnon
amplitude $v_m$ has the same form (\ref{eq:v-loc-infty}) as for
the local modes, but the function $u_m$ shows an oscillating
behaviour like
\begin{equation} \label{eq:u-oscillating}
u_m(r)\propto J_{|m+1|}(kr) + \sigma_m(k) Y_{|m+1|}(kr),
\end{equation}
where $Y_{|m+1|}$ are Neumann functions. The quantity $\sigma(k)$
derives from the soliton--magnon scattering; it can be
interpreted as the scattering amplitude.

\begin{figure}
\psfig{width=3.5in,angle=0.0,file=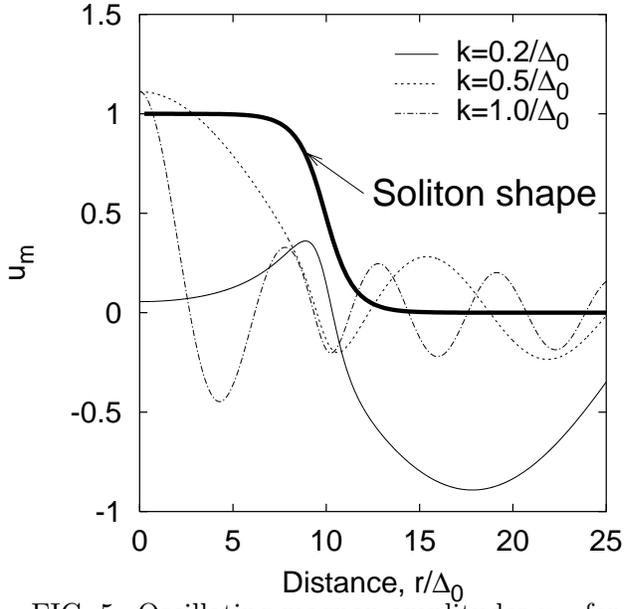}
\caption{Oscillating magnon amplitudes $u_m$ for soliton with
radius $R=10\Delta_0$ for different wave numbers $k$}
\label{fig:u_scat}
\end{figure}

We use the one--parameter shooting method to study numerically
the problem of soliton--magnon scattering, as described in
Ref.~\onlinecite{Ivanov96} for vortices in the antiferromagnet.
Choosing the shooting parameter $\varepsilon$, see
Eq.~(\ref{eq:u&v4x<<1}), we ``kill'' the growing exponent in
(\ref{eq:v-loc-infty}); as a result we have obtained a
well--pronounced exponential decay for $v_m$; the corresponding
oscillating solutions for $u_m$ differ from the asymptotical
values (\ref{eq:u-oscillating}) only in the vicinity of the
soliton core, see Fig.~\ref{fig:u_scat}. The scattering amplitude
was found from these data by comparison with the asymptotics
(\ref{eq:u-oscillating}). The results are the following:

For all modes the scattering amplitude $\sigma_m(k)$ tends to
zero as $k\to0$. In the long--wavelength limit the maximal
scattering is related to the mode with $m=-1$, for which the
behaviour of $d\sigma/dk$ looks singular. However, other
properties depend strongly on the soliton radius.

\begin{figure}
\psfig{width=3.5in,angle=0.0,file=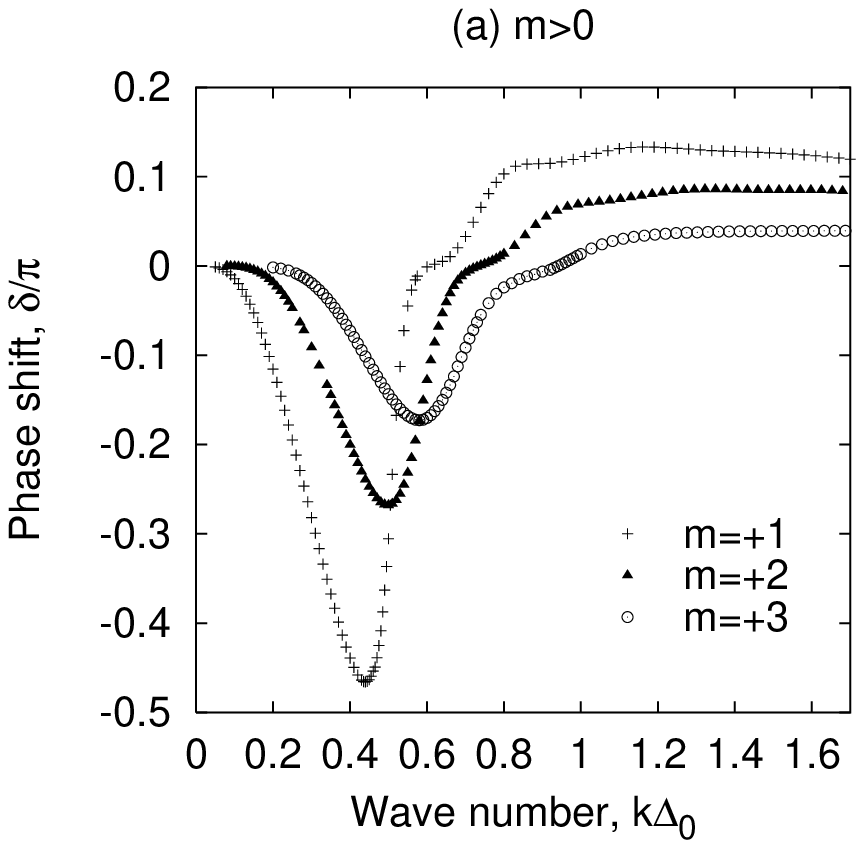}
\psfig{width=3.5in,angle=0.0,file=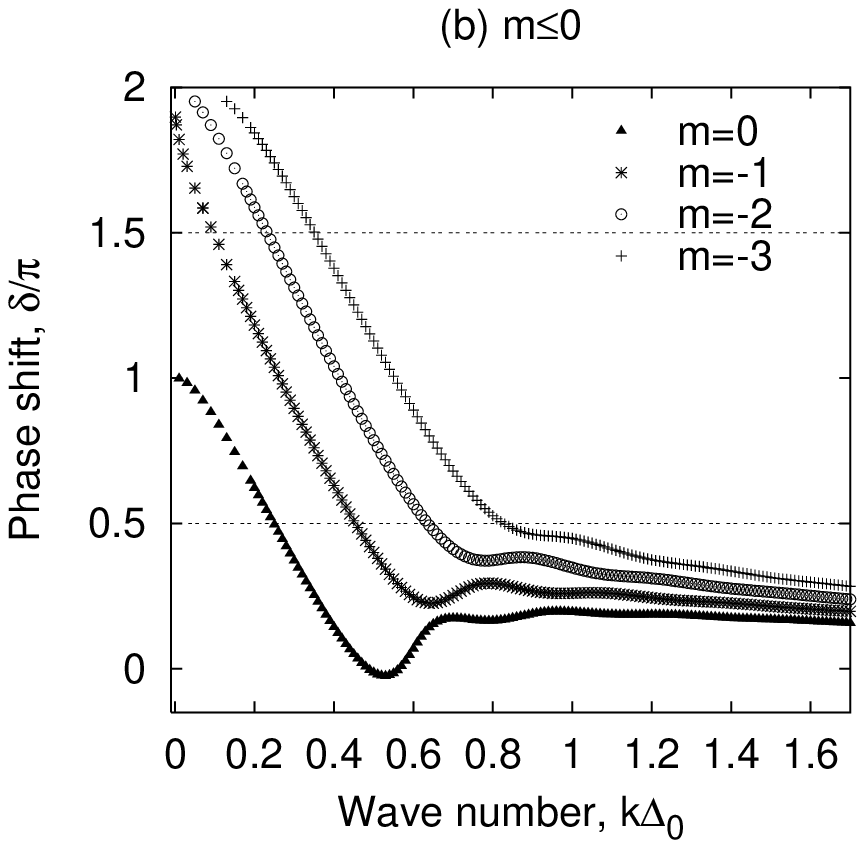}
\caption{Scattering for a large radius soliton, $R=10\Delta_0$.
Numerical data} \label{fig:delta4large_R}
\end{figure}

In the case of large $R$, the scattering amplitude $\sigma_m(k)$
is positive for all modes in the long--wavelength limit. There is
a maximum of $\sigma_m(k)$ at about $kR\sim1$ for modes with {\em
positive} $m$; but there are two poles in the scattering amplitude
for modes with {\em negative} $m$, and one pole for $m=0$.
Naturally, there is no real divergence at such a pole: the
physically observed phase of the scattering, or the phase shift
$\delta_m=-\tan \sigma_m$, varies monotonically, see
Fig.~\ref{fig:delta4large_R}. Thus, there are principally
different pictures for positive and negative $m$: the total phase
shift is
\begin{mathletters} \label{eq:delta^tot}
\begin{equation} \label{eq:delta^tot4largeR}
\delta_m^{tot} \equiv \delta(\infty)-\delta(0) = \left\{
\begin{array}{ll}
0,     & m>0\\
-\pi,  & m=0\\
-2\pi, & m<0 \\
\end{array}\right., \; R\gg\Delta_0.
\end{equation}

The scattering picture is quite different for the case of small
radius solitons. Each mode with $|m|>1$ has a single pole only,
the position $k_pR$ of which increases when $m$ grows. For the
special case of the mode with $m=+1$, the scattering amplitude is
anomalously small; for the mode with $m=-1$ there appears a
second pole with anomalously small values of $k_p$. These results
are presented in Fig.~\ref{fig:delta4small_R}. The total phase
shift is
\begin{equation} \label{eq:delta^tot4smallR}
\delta_m^{tot} = \left\{
\begin{array}{ll}
+\pi,  & m>1  \\
0,     & m=1  \\
-2\pi, & m=-1 \\
-\pi,  & m=0, m<-1
\end{array}\right.,\qquad R\ll\Delta_0.
\end{equation}
\end{mathletters}

There appears discontinuous change of the total phase shift from
(\ref{eq:delta^tot4largeR}) to (\ref{eq:delta^tot4smallR}) for
all modes with $m\neq0,-1$ for the cases of intermediate soliton
radii. This transition occurs independently for each mode at some
critical values $k_m^{\text{c}}$ of the wave--number.

\section{The scattering problem in the framework
of the pseudopotential method} %
\label{sec:pseudopotential}

For the analytical description of the scattering problem, let us
analyze the EVP qualitatively, replacing the real ``potentials''
(\ref{eq:potentials}) in the EVP (\ref{eq:u&v}) by simplified
ones, which are amenable to an exact treatment. Such an approach
is well--known in the quantum physics of solids as the
pseudopotential method. Sometimes it is useful even when the
potential is not small, see Ref.~\onlinecite{Harrison66} for
details.

\begin{figure}
\psfig{width=3.5in,angle=0.0,file=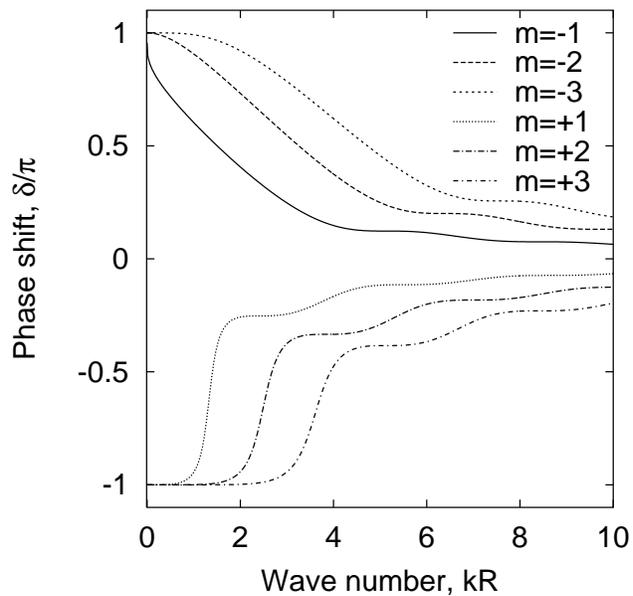}
\caption{Scattering for the ``centrifugal model''}
\label{fig:delta-centrifugal}
\end{figure}

Let us discuss the choice of the pseudopotential. The simplest
way is to describe the soliton structure in the main
approximation, using the ansatz
\begin{equation} \label{eq:delta-ansatz}
\cos\theta_0(r)\approx \text{sign }\xi, \qquad
\xi=\frac{r-R}{\Delta_0},
\end{equation}
which is a good approximation for both small and large distances.
Replacing the soliton by this configuration, which does not have
any in--plane structure ($\sin\theta_0=0$), reduces the complex
EVP (\ref{eq:u&v}) to a very simple ``centrifugal model'' with
$v_m\equiv0$:
\begin{equation} \label{eq:centrifuge}
\left(-\nabla_r^2+k^2+\frac{|m+\text{sign }\xi|^2}{r^2} \right)u_m
= 0.
\end{equation}
This model describes quasi--free magnons in each of the regions
$r<R$ and $r>R$. The only effect of the soliton--magnon
interaction is a shift of the mode indexes in comparison with the
free magnons (\ref{eq:modes_free}):
\begin{equation} \label{eq:u-cenrifugal}
u_m(r) \propto \left\{
\begin{array}{ll}
J_{|m-1|}(kr), & r<R, \\
J_{|m+1|}(kr) + \sigma_m(k) Y_{|m+1|}(kr), &r>R,
\end{array}\right.
\end{equation}
The usual matching condition for these solutions has the form
\begin{equation} \label{eq:joining-centrifugal}
\left[\frac{u'}{u}\right]_R = 0,
\end{equation}
where $\left[\dots\right]_R\equiv (\dots)\bigr|_{R+0} -
(\dots)\bigr|_{R-0}$. Calculations lead to the scattering
amplitude in the form:
\begin{equation} \label{eq:sigma-centrifugal}
\sigma(\varkappa) = \frac{ J_{|m-1|}^{\prime}(\varkappa)\cdot
J_{|m+1|}(\varkappa)- J_{|m+1|}^{\prime}(\varkappa)\cdot
J_{|m-1|}(\varkappa)} { J_{|m-1|}(\varkappa)\cdot
Y_{|m+1|}^{\prime}(\varkappa)- J_{|m-1|}^{\prime}(\varkappa)\cdot
Y_{|m+1|}(\varkappa)},
\end{equation}
where $\varkappa=kR$, see Fig.~\ref{fig:delta-centrifugal}. The
total phase shift in this model
\begin{equation} \label{eq:delta^tot4centrifugal}
\delta_m^{tot}=\pi\cdot \text{sign }m.
\end{equation}

\subsection{Small radius solitons, $\boldsymbol{R\ll\Delta_0}$}

The simple ``centrifugal model'' with the results
(\ref{eq:sigma-centrifugal}) explains qualitatively the
scattering data for the small radius solitons, except for modes
with $m=0,\pm1$, cf. (\ref{eq:delta^tot4smallR}). The cause for
these exceptions  is the existence of truly local modes
(``zero--ones'' for $m=0,+1$, and another one with non--zero
frequency for $m=-1$); while such modes are absent for the other
$m$. The existence of a local mode diminishes the total phase
shift by $\pi$ (and decreases the number of states in the
continuum spectrum by $1$, due to the conserved total number of
magnon states). Such a result is well--known in 1d, see e.g.
Ref.~\onlinecite{Ivanov95e}. Except for these cases, one can
expect for the scattering peculiarities at $k\sim
1/R>>1/\Delta_0$. In this range the approximation by an isotropic
magnet and BP--solitons is valid. The scattering problem for the
isotropic magnets was studied in detail in
Ref.~\onlinecite{Ivanov99}. Over a wide range of wave--numbers,
$1/\Delta_0\ll k \ll 1/R$, the asymptotics for the
Belavin--Polyakov soliton are valid,\cite{Ivanov99}
\begin{mathletters} \label{eq:sigma4BP}
\begin{eqnarray}
\label{eq:sigma4BP(m)}
\sigma_{m\neq-1,0}(k) &=& -\frac{\pi(kR)^2}{2|m|(m+1)},\\
\label{eq:sigma4BP(m=0)}
\sigma_{m=0}(k) &=& \frac{\pi}{2\ln\left( 1/kR\right)},\\
\label{eq:sigma4BP(m=-1)} \sigma_{m=-1}(k) &=& \pi(kR)^2\ln\left(
1/kR\right).
\end{eqnarray}
For larger wave--numbers, $k\gg|m|/R$, the scattering amplitude
has a general, so--called, eikonal dependence,
$\sigma_m\propto1/kR$, \cite{Ivanov99}
\begin{equation} \label{eq:sigma4BP(kR>>1)}
\sigma_m(k) = \frac{\pi(m-1)}{kR}.
\end{equation}
\end{mathletters}

\begin{figure}
\psfig{width=3.5in,angle=0.0,file=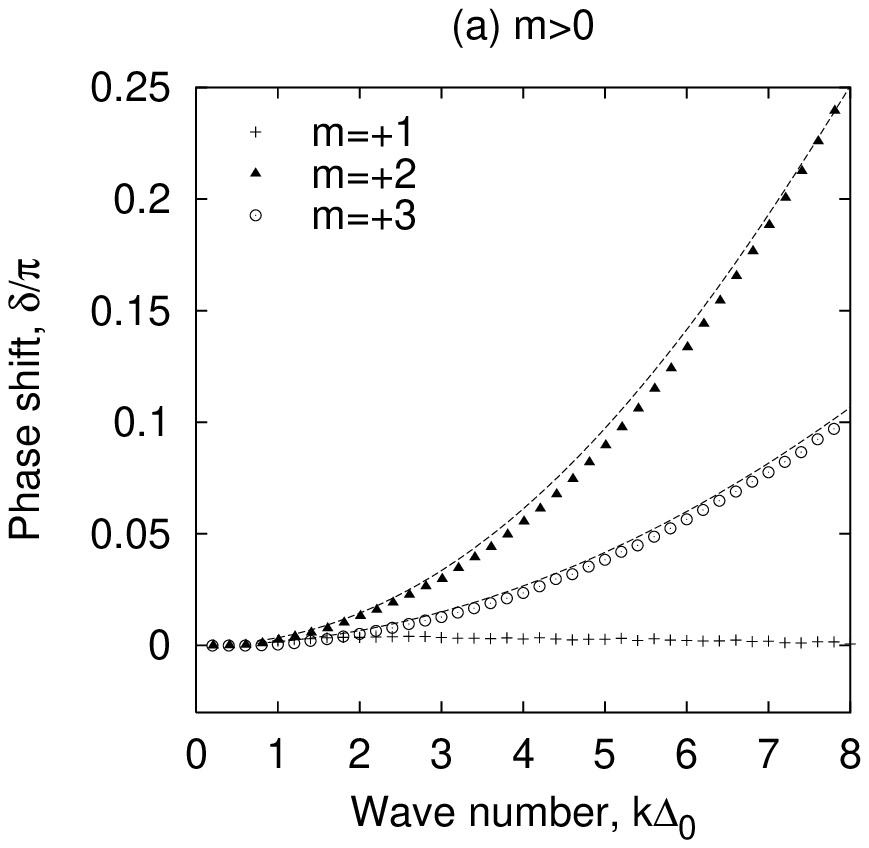}
\psfig{width=3.5in,angle=0.0,file=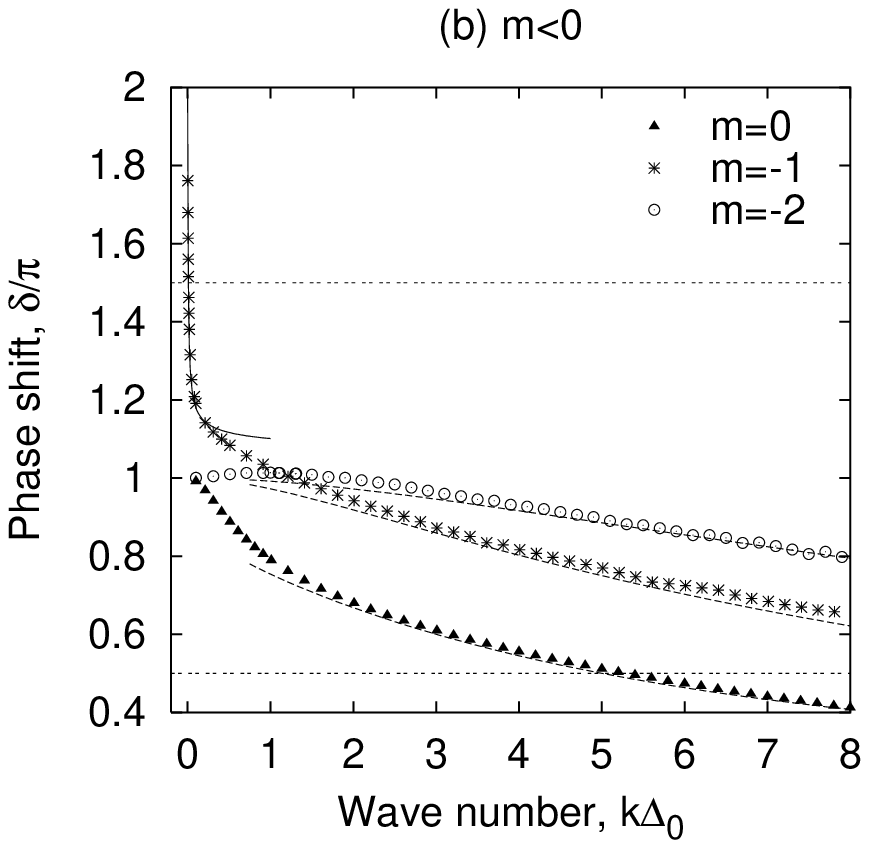}
\caption{Scattering for a small radius soliton, $R=0.2\Delta_0$.
Solid line: analytical asymptotics for $k\ll1/\Delta_0$ for the
mode $m=-1$ from Eqs.~(\ref{eq:sigma4m=-1(kR<<1)}); dashed lines:
analytical asymptotics for $k\gg 1/\Delta_0$ from the
Eqs.~(\ref{eq:sigma4BP}); symbols: numerical data}
\label{fig:delta4small_R}
\end{figure}

Our numerical data in this region are very close to the
asymptotical behaviour (\ref{eq:sigma4BP}), see
Fig.~\ref{fig:delta4small_R}. Note that the maximal scattering in
the region $1/\Delta_0\ll k \ll 1/R$ occurs for the rotational
mode with $m=0$, whose asymptotics (\ref{eq:sigma4BP(m=0)}) has a
singularity for $k\to0$. There is one pole only for each mode;
its position corresponds to the pole in the Belavin--Polyakov
solution, $k_p\sim|m|/R$.\cite{Ivanov99}

Let us consider the local modes, for which the total phase shift
is diminished by $\pi$ in comparison to the ``centrifugal model''
(\ref{eq:delta^tot4centrifugal}); it is equal to $0$, $-\pi$,
$-2\pi$ for the cases $m=+1$, $0$, $-1$, respectively.

There is one pole only for the simplest case of the mode with
$m=0$. The scattering amplitude in the range $1/\Delta_0\ll k \ll
1/R$ is described by the formula (\ref{eq:sigma4BP(m=0)}). The
position of the pole is determined by the BP--result
$k_p=1/R$.\cite{Ivanov99} The reason for such a nice agreement
with the model of the isotropic magnet will be explained below in
the text.

\begin{figure}
\psfig{width=3.5in,angle=0.0,file=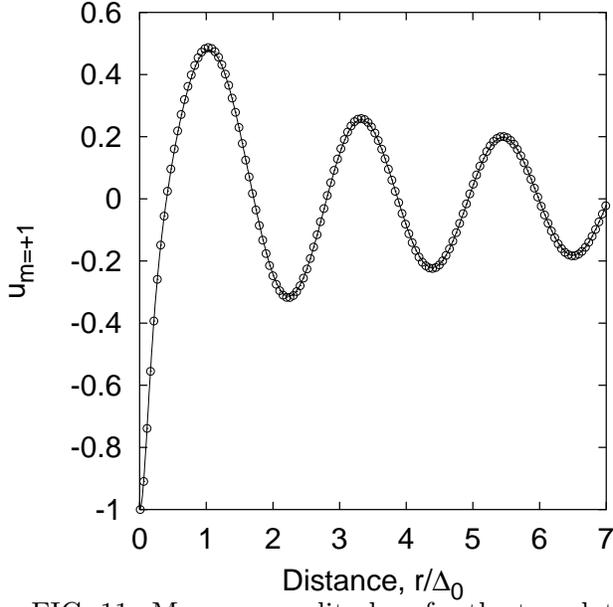}
\caption{Magnon amplitude $u$ for the translational mode with
$m=+1$ ($R=0.2\Delta_0$, $k\Delta_0=3$). Lines: exact analytical
solution (\ref{eq:u4BP}) for the isotropic magnet; symbols:
numerical data for the EA FM} %
\label{fig:BP-amplitudes}
\end{figure}

There is no pole for the translational mode with $m=+1$. This is
due to a unique property of the isotropic model, where such a
mode does not scatter at all;\cite{Ivanov99} the magnon amplitude
in this limit has the form
\begin{equation} \label{eq:u4BP}
u_{m=+1}(r) = J_2(kr) -
\frac{2}{kr}\cdot\frac{J_1(kr)}{(r/R)^2+1},
\end{equation}
which agrees with our numerical data, see
Fig.~\ref{fig:BP-amplitudes}. This unique property of
reflectionlessness is caused by the high internal symmetry of the
problem.

Fig.~\ref{fig:delta4small_R}(a) shows that such a property is
approximately valid for the small radius soliton in the
anisotropic magnet, too: the scattering amplitude is anomalously
small, so the total phase shift is absent. Another effect of this
symmetry in the isotropic magnet is that the scattering problem
for the BP--model can be reduced to a simplified one with
effectively smaller scattering. Such a ``reduced model'' has no
local (and quasi--local) modes; it is described by shifted
indexes, $m\to m-1$.\cite{Ivanov99} It is natural to use the
``centrifugal'' approximation for such a reduced model
(\ref{eq:sigma-centrifugal}).  As a result, $\sigma_{m=+1}=0$ in
the framework of the same ``centrifugal model''; for the other
modes, Eq.~(\ref{eq:delta^tot4centrifugal}) leads to the total
phase shift $\delta_m^{tot}=\pi\cdot \text{sign}(m-1)$. Taking
into account that $\text{sign }0=0$, one can see that such a
simple dependence for $\delta_m^{tot}$ describes all modes, except
of $m=-1$. Its behaviour is most interesting.

For the mode with $m=-1$ there are two poles, while it should be
one pole only for the Belavin--Polyakov solitons. Indeed one of
the poles lies in the region where the Belavin--Polyakov
approximation is valid; its value,
$k_p\approx2.79/R\gg1/\Delta_0$, coincides exactly with the
Belavin--Polyakov data.\cite{Ivanov99} The second pole lies in
the region $k\ll1/\Delta_0$, where the approximation by the
isotropic magnet is not applicable. This pole is caused by the
presence of the local mode, the typical wave--number of such a
mechanism is $k_p\sim\varkappa_1^\pm\ll1/\Delta_0$. The use of
``reduced'' or ``centrifugal''  models is certainly wrong for
this case. We will discuss the scattering amplitude in this
region later, on the basis of a more general pseudopotential
model, see Eq.~(\ref{eq:sigma4m=-1(kR<<1)}).

\subsection{Large radius solitons, $\boldsymbol{R\gg\Delta_0}$.
Refinement of the pseudopotential model}

Let us consider the case of large radius solitons,
$R\gg\Delta_0$. Here exist local modes for each $m$. Therefore the
simple ``centrifugal model'' as well as other simple models
without a discrete spectrum seem to be wrong. This can be
explained by the fact that in the very simple model
(\ref{eq:centrifuge}) we omit the ``coupling potential''; the
real potential $W$ tends to zero at the origin and at infinity,
but has maxima at some finite distances.

Thus we need a more general form of the pseudopotentials which
accounts for the potentials $U_0$, $V$, and $W$. To describe the
soliton shape, we can again use the expression
(\ref{eq:delta-ansatz}) for the out--of--plane soliton
configuration. However, we need to consider the in--plane
signature, too. According to (\ref{eq:delta-ansatz}), it could be
described as $$\sin^2\theta_0(r)\longrightarrow \delta(\xi).$$
Therefore we can suppose that the ``coupling potential'' $W(r)$
is nonzero only at some distance around the soliton radius and
approximately replace it by a $\delta$--potential as follows,
\begin{mathletters} \label{eq:delta-potentials}
\begin{equation} \label{eq:W-delta}
W(r)\longrightarrow \frac{c_1}{\Delta_0^2}\delta(\xi).
\end{equation}

The same assumptions lead to simple expressions for the other
potentials in the EVP (\ref{eq:u&v}),
\begin{eqnarray} \label{eq:V-delta}
&& U_0(r)+\frac{m^2}{r^2} \pm V(r) \nonumber \\
&& \longrightarrow \frac{1 -(\Omega/\omega_0)\cdot\text{sign
}\xi}{\Delta_0^2}+ \frac{|m\pm\text{sign}\xi|^2}{r^2} +
\frac{c_2}{\Delta_0^2}\delta(\xi).
\end{eqnarray}
\end{mathletters}
Here $c_1$ and $c_2$ are trial parameters.

We choose this pseudopotential model (\ref{eq:delta-potentials})
out of the following reasons: (i) it allows an exact analytical
solution; (ii) it guarantees the correct asymptotic behaviour of
the solutions; (iii) it offers a possibility to identify the
parameters of the pseudopotential from the discrete spectrum
(\ref{eq:omega_m_cases}).

Let us consider the discrete spectrum in the framework of the
pseudopotential (\ref{eq:delta-potentials}). Comparison with the
eigenfrequencies (\ref{eq:omega_m_cases}) provides a possibility
to determine all trial parameters $c_k$. For the discrete part of
the spectrum,  the EVP (\ref{eq:u&v}) with potentials
(\ref{eq:delta-potentials}) has solutions like
\begin{mathletters} \label{eq:u&v-delta-pot}
\begin{eqnarray} \label{eq:u-delta-pot}
&&u_m(r) \propto \left\{
\begin{array}{ll}
I_{|m-1|}(\varkappa_2^-r), & r<R \\
K_{|m+1|}(\varkappa_1^-r), & r>R
\end{array}\right.,
\\
\label{eq:v-delta-pot} %
&& v_m(r) \propto \left\{
\begin{array}{ll}
I_{|m+1|}(\varkappa_2^+r), & r<R \\
K_{|m-1|}(\varkappa_1^+r), & r>R
\end{array}\right.,\\
&&
\varkappa_1^\pm=\sqrt{\left(\omega_0-\Omega\pm{\omega}\right)/D},\quad
\varkappa_2^\pm=\sqrt{\left(\omega_0+\Omega\pm{\omega}\right)/D}.\nonumber
\end{eqnarray}
\end{mathletters}

Now it is possible to calculate the eigenspectrum by matching the
solutions (\ref{eq:u-delta-pot}) and (\ref{eq:v-delta-pot}) at
the distance $R$,
\begin{mathletters} \label{eq:u&v-joining}
\begin{eqnarray}\label{eq:u&v-joining(1)}
&&\Bigl[u^\prime\Bigr]_R = \frac{c_2}{\Delta_0}\cdot u\Bigr|_R-
\frac{c_1}{\Delta_0}\cdot v\Bigr|_R, \\ %
\label{eq:u&v-joining(2)} %
&&\Bigl[v^\prime\Bigr]_R = \frac{c_2}{\Delta_0}\cdot
v\Bigr|_R-\frac{c_1}{\Delta_0}\cdot u\Bigr|_R.
\end{eqnarray}
\end{mathletters}
Calculations lead to the eigenfrequencies in the form
(\ref{eq:omega_m}). Using the condition for the ``zero--modes''
(\ref{eq:omega=0}), the eigenfrequencies can be represented in the
form (\ref{eq:omega_m*}); the constants $c_1$, $c_2$ can be
identified by comparison with the spectrum (\ref{eq:omega_m*}),
\begin{equation} \label{eq:c1-c2}
c_1 = \frac12\left(\beta-\frac{\alpha^2}{\beta}\right), \; c_2 =
-2+c_1+m\alpha+m^2\beta-\frac{\Delta_0^2}{4R^2}.
\end{equation}
The parameters $\alpha$ and $\beta$ were calculated by the
variational approach, $\alpha\propto1/R^3$, $\beta\propto1/R^2$,
see Eq.~(\ref{eq:alpha&beta}).

Let us discuss the scattering problem. Solving the EVP
(\ref{eq:u&v}) with pseudopotentials (\ref{eq:delta-potentials}),
we obtain for the continuum spectrum,
${\omega}>\omega_{\text{gap}}$, the following solutions. The
function $u$ for $r<R$ has the form
\begin{equation} \label{eq:u-delta-pot-scat}
u(r) \propto \left\{
\begin{array}{ll}
I_{|m-1|}(\varkappa_2^-r), & \text{when $k<\sqrt{2\Omega/D}$}, \\
J_{|m-1|}(k_2^-r),         & \text{when $k>\sqrt{2\Omega/D}$},
\end{array}\right.
\end{equation}
where $k_2^- = \sqrt{k^2-2\Omega/D}$. At large distances, $r>R$,
it has the usual oscillating form (\ref{eq:u-oscillating}). The
function $v$ has the localized form (\ref{eq:v-delta-pot}). One
can find the scattering amplitude by matching the solutions
(\ref{eq:u-oscillating}), (\ref{eq:u-delta-pot-scat}), and
(\ref{eq:v-delta-pot}) at the distance $R$, using the condition
(\ref{eq:u&v-joining}):
\begin{equation} \label{eq:sigma-pseudo}
\sigma_m(k) = \frac{\left[{\frak C}_m(k)+{\frak F}_m(k)\right]
J_{|m+1|}(kR)- k\Delta_0 J_{|m+1|}^{\prime}(kR)}{k\Delta_0
Y_{|m+1|}^{\prime}(kR)-\left[{\frak C}_m(k)+{\frak F}_m(k)\right]
Y_{|m+1|}(kR)},
\end{equation}
here
\begin{eqnarray*}
{\frak C}_m(k)&=& c_2-c_1^2\cdot\Biggl(c_2+
\varkappa_2^+\Delta_0\cdot
\frac{I_{|m+1|}^{\prime}(\varkappa_2^+R)}{I_{|m+1|}(\varkappa_2^+R)}
\\ &-&\varkappa_1^+\Delta_0\cdot
\frac{K_{|m-1|}^{\prime}(\varkappa_1^+R)}{K_{|m-1|}(\varkappa_1^+R)}
\Biggr)^{-1},\\
{\frak F}_m(k) &=& \left\{
\begin{array}{ll}
\varkappa_2^- \Delta_0\cdot{\displaystyle
\frac{I_{|m-1|}^{\prime}( \varkappa_2^-R)}{I_{|m-
1|}(\varkappa_2^-R)}},
& \text{$k<\sqrt{2\Omega/D}$},\\
k_2^-\Delta_0\cdot{\displaystyle\frac{J_{|m-1|}^{\prime}(
k_2^-R)}{J_{|m-1|}(k_2^-R)}}, & \text{$k>\sqrt{2\Omega/D}$}.
\end{array}\right.
\end{eqnarray*}

It is easy to analyze the expression for the scattering amplitude
(\ref{eq:sigma-pseudo}) in some limiting cases. In the
long--wavelength limit, $k\ll1/R\ll1/\Delta_0$, one can use the
asymptotics for the cylindrical functions $J(z)$, $Y(z)$ at
$z\ll1$ and, at the same time, the asymptotics for $I(z)$, $K(z)$
for  $z\gg1$. Simple calculations show that the scattering
intensity is maximal for the mode with $m=-1$,
\begin{mathletters} \label{eq:sigma4kR<=1}
\begin{equation} \label{eq:sigma4m=-1(kR<<1)}
\sigma_{m=-1}(k)=\frac \pi {2\ln \left( 1/kR\right) }, \quad
kR\ll1,
\end{equation}
which agrees with the numerical data, see
Fig.~\ref{fig:delta4large_R}(b). Note, that the
Eq.~(\ref{eq:sigma4m=-1(kR<<1)}) was obtained using one condition
only, $kR\to0$. Therefore it is valid for the case of small $R$,
too, explaining the existence of the second pole for the mode
$m=-1$ both for large and small radii.

For the other values of $m$, we can restore the general dependence
of the amplitude of the magnon scattering on the large radius
soliton, $R\gg\Delta_0$:
\begin{equation} \label{eq:sigma4m(kR<<1)}
\sigma_{m\neq-1}(k) = \frac{\pi\left(kR/2\right)^{2|m+1|}}{
|m+1|!(|m+1|-1)!} , \quad kR\ll1.
\end{equation}
The scattering amplitude, according to
Eq.~(\ref{eq:sigma4m(kR<<1)}), is positive for all modes in the
long--wavelength limit.

In the range $1/R\ll k \ll 1/\sqrt{R\Delta_0}$ the asymptotics
for $J(z)$, $Y(z)$ for $z\gg1$ are valid, which in the main
approximation for $\Delta_0/R$ leads to a linear dependence of
the scattering phase shift on $kR$,
\begin{equation} \label{eq:delta4m(kR>>1)}
\delta_m(k) \approx \frac{|m+1|\pi}{2}-\frac{\pi}{4}-kR, \quad
1\ll kR\ll\sqrt{\frac{R}{\Delta_0}}.
\end{equation}
\end{mathletters}
For this range of $k$, all curves look like a family of parallel
lines with equal distance of about $\pi/2$. Note that the real
values of the trial parameters $c_1$ and $c_2$ lead to the
corrections in Eqs.~(\ref{eq:sigma4kR<=1}) like $\Delta_0/R$.

\begin{figure}
\psfig{width=3.5in,angle=0.0,file=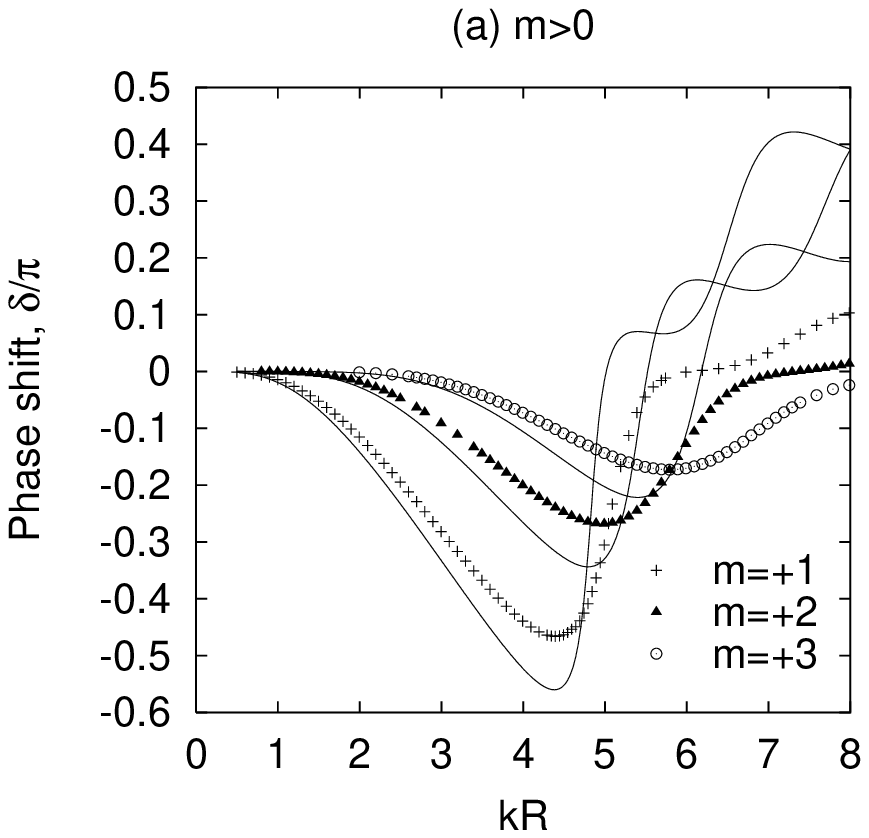}
\psfig{width=3.5in,angle=0.0,file=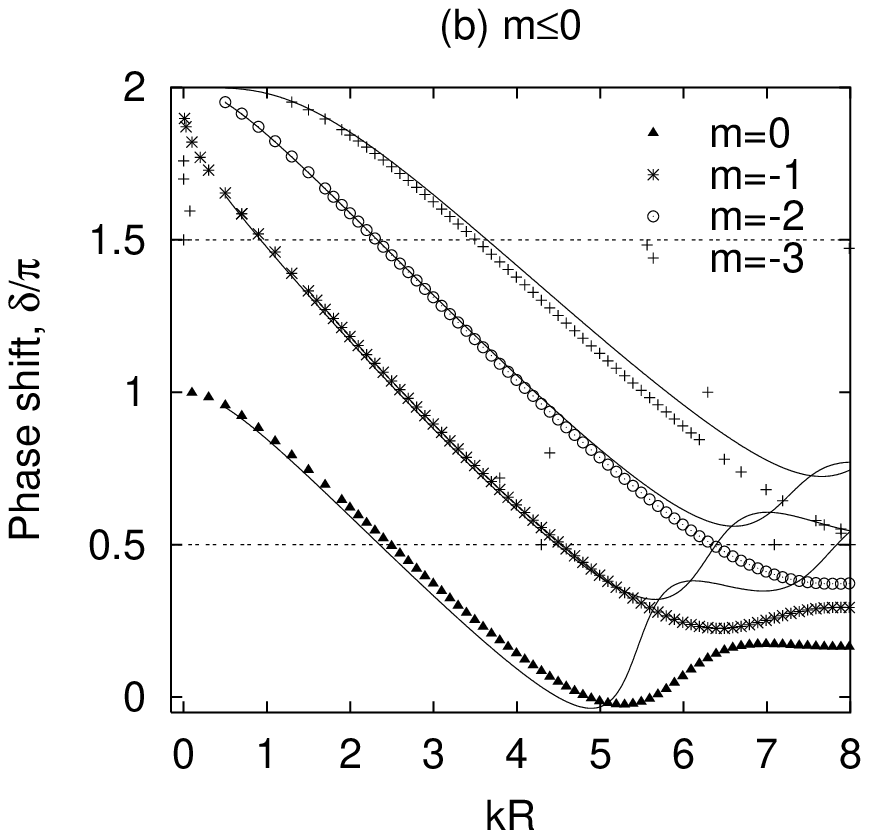}
\caption{Scattering amplitude $\delta_m$ for large radius soliton
($R=10\Delta_0$) vs. $kR$ in the range $k\Delta_0<0.8$. Lines:
analytical solution from (\ref{eq:sigma-pseudo}); symbols:
numerical data} %
\label{fig:sigma-pseudo}
\end{figure}

Naturally the pseudopotential method cannot explain the scattering
results for $k\gtrsim1/\Delta_0$, where the influence of truly
local modes and the differences between the real potential and the
$\delta$--function are dominant. Nevertheless, the method gives
quite good results even for the case $k\lesssim1/\Delta_0$, where
the linear asymptotics does not work. Corresponding curves are
plotted in Fig.~\ref{fig:sigma-pseudo} using the general
dependence (\ref{eq:sigma-pseudo}) with trial parameters $c_1$
and $c_2$. Fitting of these parameters gives a result in the
zeroth--order approach for $\Delta_0/R$: $c_1\approx0$,
$c_2\approx0.4$ instead of $c_1\approx0$, $c_2\approx-2$,
according to (\ref{eq:c1-c2}). The difference is caused by the
limitations of the model (\ref{eq:delta-potentials}). For
example, the above mentioned model leads to an increase of the
scattering phase when $k$ increases; the total phase shift for
this model is $\delta_m^{tot}=\text{sign }m\cdot\pi$, as for the
``centrifugal model'', see (\ref{eq:delta^tot4centrifugal}), i.e.
the qualitative properties of these models are equivalent. The
appearance of a local mode diminishes the total phase shift by
$\pi$. As a result, there is no total phase shift for all modes
with positive $m$; but $\delta^{tot}=-2\pi$ for all modes with
$m<0$.

The scattering picture in the short--wavelength limit,
$k\gg|m|/R$, gives the general, so--called eikonal dependence
$\sigma_m\approx A_m/k$. However, in contrast to the case of the
isotropic magnet (\ref{eq:sigma4BP(kR>>1)}), the scattering
amplitude has the same sign $A_m<0$ for all modes due to the
presence of the effective attractive potential, which causes the
existence of local modes. When the soliton radius decreases and
the local mode disappears, the sign of $\sigma_m$ changes; the
result becomes similar to (\ref{eq:sigma4BP(kR>>1)}).

\begin{figure}
\psfig{width=3.5in,angle=0.0,file=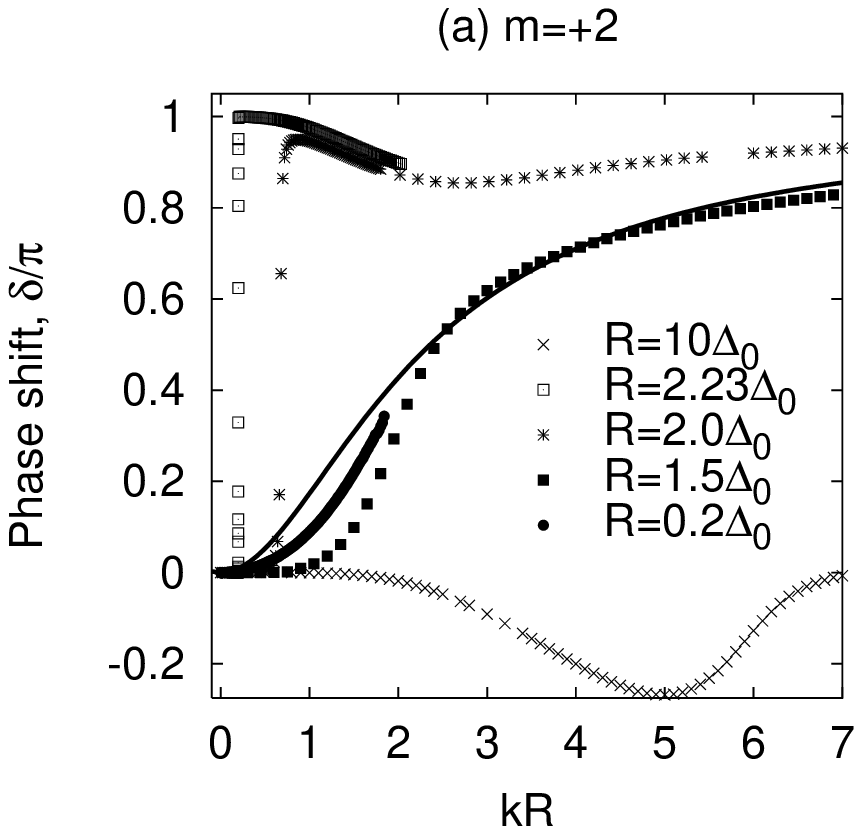}
\psfig{width=3.5in,angle=0.0,file=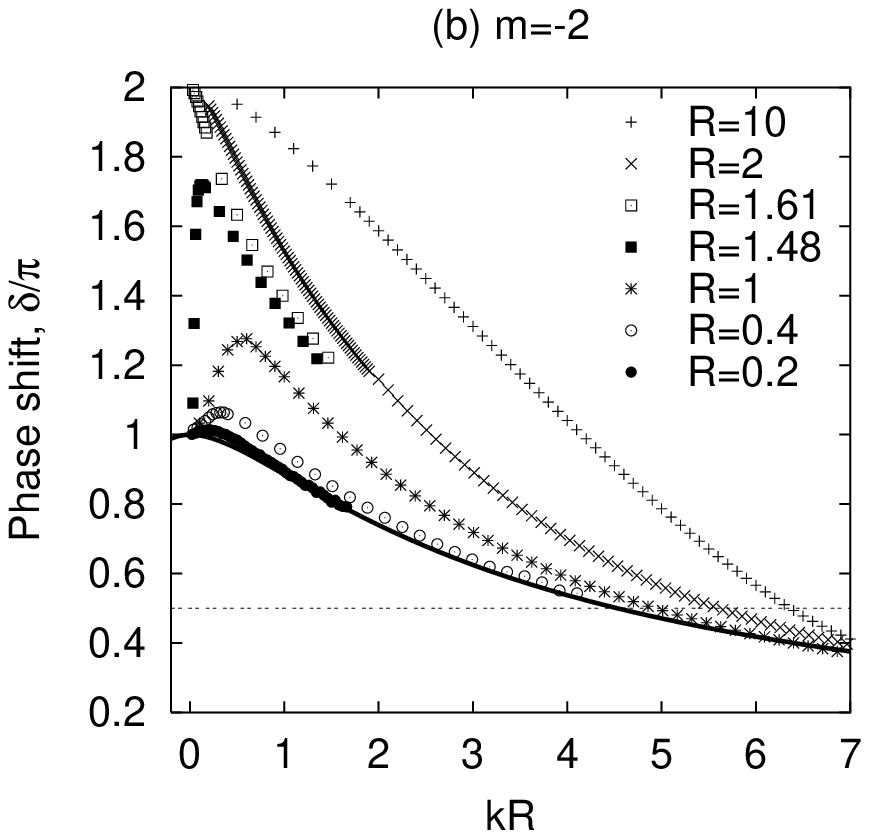}
\caption{Bifurcations for the scattering data $\delta_m$ vs.
$kR$: (a) critical soliton radius
$R_{m=+2}^{\text{c}}=2.24\Delta_0$, (b)
$R_{m=-2}^{\text{c}}=1.52\Delta_0$. Solid line corresponds to the
Belavin--Polyakov asymptotics
(\ref{eq:sigma4BP}).} %
\label{fig:delta_m_2.dif-R} %
\end{figure}

Thus, we explain the particularities in the scattering picture of
EA magnets, in comparison with the isotropic case, by the
influence of local modes. To examine this influence, let us
consider the scattering data $\sigma_m$ for the modes $m=+2$ and
$m=-2$ with different soliton radii $R$ as functions of the
parameter $kR$, i.e. $\sigma=\sigma_m(kR)$; this gives a
possibility to compare the scattering data with the isotropic
model, see Fig.~\ref{fig:delta_m_2.dif-R}. For the small radii the
scattering picture is similar to the BP--case (solid lines in
Fig.~\ref{fig:delta_m_2.dif-R}). When the soliton radius
increases, the scattering phase jumps up by $\pi$, decreasing the
total phase shift. Such a bifurcation takes place for the mode
with $m=+2$ at $R=2.24\Delta_0$, and for the mode with $m=-1$ at
$R=1.52\Delta_0$. These values agree with the crossover points
$R^{\text{c}}_m$, see Table~\ref{table:Rc}. Thus, it is natural
to suppose that the bifurcation is connected with the appearance
of the quasi--local modes inside the continuum spectrum for
$R>R^{\text{c}}_m$. Simple calculations show that the positions
$k_p$ of the poles correspond to the frequencies of quasi--local
modes, which can be estimated by Eq.~(\ref{eq:omega_m(R>>1)}).

\section{Translational modes and description of the soliton dynamics}
\label{sec:eq-mot}

The foregoing analysis of the spectrum for the modes with $|m|=1$
can be used to describe the translational motion of the soliton.
Such an approach was first proposed in
Ref.~\onlinecite{Wysin96a}, where an effective mass of the vortex
in a 2d EP magnet was calculated from numerical data for the
eigenmode spectrum of the finite size magnet. An important
progress for the vortices in FM was achieved in
Ref.~\onlinecite{Ivanov98}; the specific structure of the
spectrum of modes with $|m|=1$ (namely the existence of a
Goldstone mode with $\omega\propto1/L^2$, and a number of
doublets with mean frequency $\overline{\omega}\propto1/L$ and a
small splitting) leads to non--Newtonian equations of motion with
$3^{\text{rd}}$, $5^{\text{th}}$, \dots\ time derivatives. Such
equations were derived and discussed earlier in
Ref.~\onlinecite{Mertens97}, using a phenomenological approach,
for a review see Ref.~\onlinecite{Mertens99}. The analytical
calculation\cite{Ivanov98} verified their adequacy; besides, it
allows to calculate the coefficients in the equations of motion
with an error of only about $0.8\%$. Thus, it is possible to
describe in this way such unusual properties as the appearance of
the non--Newtonian equations of motion.

The spectrum of magnon modes with $|m|=1$ has a different
structure for the isotropic FM. First of all, there are no
doublets;\cite{Ivanov99} for an approximate description of the
dynamics only two modes are sufficient: the translational
Goldstone mode with dependence $\omega\propto1/L^2$ and another
mode with $m=-1$. Therefore, the effective equations of motion
have a Newtonian form with an effective mass, which diverges as
$L^2$, see Ref.~\onlinecite{Ivanov99}. Such behaviour agrees with
the direct calculations done by Zaspel.\cite{Zaspel93}

Let us note that all magnets mentioned above, i.e. EP and
isotropic systems, have a gapless dispersion law and, as a
result, a strong interaction of the soliton with the boundaries of
the magnet; the interaction force is proportional to $1/L$ for
the EP vortices\cite{Ivanov98} and $1/L^2$ for the
Belavin--Polyakov solitons.\cite{Ivanov99} The absence of a gap
manifests itself in the particularities of the dynamical
coefficients (the effective mass $M$ for the BP--soliton in the
isotropic FM\cite{Ivanov99} and the coefficient ${G}_3$ in the
term with $\left[\boldsymbol{e}_z\times\partial^3{\boldsymbol{X}}/
\partial t^3\right]$ for the vortex in the EP FM\cite{Mertens97}).
This can be explained as follows. The existence of the mass $M$
and the gyroscopical force $G\left[\boldsymbol{e}_z\times
\partial{\boldsymbol{X}}/\partial t\right]$ leads to a finite
frequency of the Larmor precession of the soliton. For a gapless
dispersion law, this frequency lies in the continuous spectrum.
Thus, the Larmor motion of the vortex leads to the generation of
magnons. For a finite--size system without dissipation, such
magnons are distributed through the whole magnet; the Larmor
dynamics of the soliton centre is strongly coupled with a
``magnon cloud'', having a scale of about the system size.
Therefore, the position of the soliton centre, which can be
determined as a point with $m_z=-1$ (BP--soliton), or $m_z=\pm1$
(vortex), in fact plays the role of a collective variable which
governs the motion of the ``magnon cloud''. Then it is not
surprising that the corresponding equations of motion are
nonlocal, leading to the divergence of the dynamic coefficients
as $L\to\infty$.

The situation is quite different for the case of the EA FM. Our
numerical analysis and analytical calculations show that the
picture of doublets for local modes with $|m|$ and $-|m|$ is valid
for large $R$, see Sec.~\ref{sec:local}. On the other hand, the
eigenfrequencies of these modes lie in the gap for the large
soliton radii, $\omega<\omega_{\text{gap}}$, so one can expect the
existence of a finite mass in the effective equations of the
soliton motion. Therefore for the phenomenological description of
the soliton dynamics we can use the $2^{\text{nd}}$--order
differential equation
\begin{equation} \label{eq:EqMot}
M\frac{\partial^2{\boldsymbol{X}}}{\partial t^2} -
G\left[\boldsymbol{e}_z\times \frac{\partial{
\boldsymbol{X}}}{\partial t}\right] = \boldsymbol{F}_e
\left(\boldsymbol{X}\right),
\end{equation}
which corresponds to the picture with the lowest doublet of local
modes. Note that further generalizations for the finite size
magnet take into account the next doublets of quasi--continuous
modes. Here the hierarchy of the effective equations of motion
containing only even--order time derivatives manifests itself. In
Eq.~(\ref{eq:EqMot}) $\boldsymbol{X}$ describes the position of
the soliton, $M$ is the mass coefficient, $G$ the
gyrocoefficient, $\boldsymbol{e}_z$ the unit vector along the
easy ($z$) axis, and $\boldsymbol{F}_e$ the external force acting
on the soliton due to the boundary or other solitons. Assuming
that deviations of the soliton from the equilibrium position are
small, $\boldsymbol{F}_e \approx-\alpha\boldsymbol{X}$, the
effective equation of motion (\ref{eq:EqMot}) can be solved by a
harmonic ansatz, which leads to
\begin{equation} \label{eq:omega12+Fe}
\omega_{0,1}=\frac{G}{2M}\pm \sqrt{\frac{G^2}{4M^2}+
\frac{\alpha}{M}}.
\end{equation}
For the localized solitons, the interaction between soliton and
boundary (or between two solitons) has an exponential decay as
$\exp(-L/r_0)$, $r_0=\sqrt{D/\omega_{\text{gap}}}$, see
Eq.~(\ref{eq:theta_r<<1}). Thus $\alpha/M$ can be neglected for a
large system size $L$, so the frequencies (\ref{eq:omega12+Fe})
have the form:
\begin{equation} \label{eq:omega12}
\omega_0=0,\qquad \omega_1 = G/M.
\end{equation}
The zero frequency $\omega_0$ corresponds to the Goldstone mode
(position shift of the soliton), in which all spins rotate with
the frequency of the soliton precession $\Omega$ in the
laboratory frame. The presence of the rotating frame frequencies
in the phenomenological description of the soliton dynamics has a
simple explanation:

The frequencies (\ref{eq:omega12}) describe, in essence, the
motion of the soliton centre. Let us discuss their link with the
frequencies of small oscillations, see
(\ref{eq:vartheta&mu-via-u&v}). It is easy to write down
expressions for the magnetization components in the laboratory
coordinate frames:
\begin{eqnarray*}
m_x+im_y&=&\Bigl[\sin\theta_0+\cos\theta_0 \sum_{\alpha} \left(
u_\alpha + v_\alpha \right)\cos(m\chi+\tilde{\omega}_\alpha t)\\
&+& i\sum_{\alpha} \left( u_\alpha - v_\alpha
\right)\sin(m\chi+\tilde{\omega}_\alpha t)\Bigr]\cdot
e^{iq\chi+i\Omega t},\\ %
m_z &=& \cos\theta_0 - \sin\theta_0 \sum_{\alpha} \left( u_\alpha
+ v_\alpha \right)\cos(m\chi+\tilde{\omega}_\alpha t).
\end{eqnarray*}
One can see that the in--plane components have no simple frequency
dependence, including the combined values
$\tilde{\omega}\pm\Omega$. At the same time, the out--of--plane
components, $m_z$, depend on the frequency $\tilde{\omega}$ in
the rotating frame only. In our phenomenological description only
the $z$--components contain information about the position of the
soliton centre $\boldsymbol{X}$, determined by the condition
$m_z=-1$. Thus, just the frequency $\tilde{\omega}$, which is
calculated above and plotted in all figures, determines the
soliton motion.

\begin{figure}
\psfig{width=3.5in,angle=0.0,file=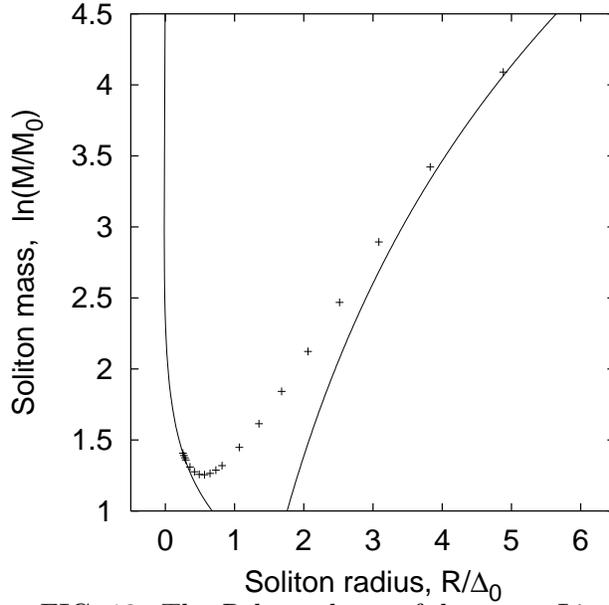} \caption{The $R$
dependence of the mass. Lines: analytical asymptotics from
Eqs.~(\ref{eq:mass}); symbols: numerical data} \label{fig:mass}
\end{figure}

The solution of the effective equation (\ref{eq:EqMot}) with the
nonzero frequency $\omega_1$ allows us to calculate the effective
mass of the soliton, $M=G/\omega_1$. The value of the
gyroconstant is well--known, $G = 4\pi q A/D$, see
Refs.~\onlinecite{Bar'yakhtar93,Ivanov95e}. As a result, the
soliton mass is finite for every soliton radius (the local mode
with $\omega_1$ exists for every $R$, see Fig.~\ref{fig:omega}).
This fact  corresponds qualitatively to the calculations in
Ref.~\onlinecite{Ivanov89}. The soliton mass we have obtained
numerically from the $\omega_1$--data is plotted in
Fig.~\ref{fig:mass} as a function of $R$.

In the limit of a large radius soliton the formula
(\ref{eq:omega_m=-1}) yields the dependence
\begin{mathletters}\label{eq:mass}
\begin{equation}\label{eq:mass4large_R}
M = M_0\left(\frac R\Delta_0\right)^3, \quad M_0=\frac{\hbar
N_2}{D},\quad R\gg \Delta_0,
\end{equation}
which agrees with the results obtained in
Ref.~\onlinecite{Ivanov89} and our numerical data, see
Fig.~\ref{fig:mass}.

In the case of small radius, $R\ll\Delta_0$, the dependence
(\ref{eq:omega_m=-1_small_R}) is valid, thus
\begin{equation}\label{eq:mass4small_R}
M = M_0\cdot\frac{2\omega_0}{\omega_0-\Omega}
  = \frac{4\pi A}{D^2}\cdot r_0^2, \; R\ll \Delta_0,
\end{equation}
\end{mathletters}
where $r_0$ was introduced in Eq.~(\ref{eq:theta_r<<1}). This
result corresponds to the one for the Belavin--Polyakov
soliton,\cite{Ivanov99} $M = 4\pi A D^{-2}\cdot L^2$ , because
here the characteristic length $r_0$ is to be replaced by the
system size $L$.

\section{Conclusion}
\label{sec:conclusion}

We have studied the magnon--soliton system in the model of the 2d
Heisenberg easy--axis ferromagnet. Combining numerical and
analytical methods, we have obtained complete results about bound
and scattering magnon states in the system. A rich spectrum of
truly local modes was found along with bifurcations of these modes
with the change of the soliton radius. For the modes with higher
azimuthal numbers $m$, we have verified the picture of doublets
with small splitting for large soliton radii $R$. The spectrum
changes strongly with decreasing of $R$: all modes with $|m|>1$
``leave'' the region of the discrete spectrum, transforming to
quasi--local modes. Such modes could be observed experimentally
by soliton magnetic resonance as was done for 1d
solitons.\cite{Boucher87}

The influence of truly local modes is important for the scattering
problem, leading to the bifurcations of the phase shift. As a
result, when $R<1.52\Delta_0$, the scattering picture is
qualitatively the same as for the isotropic magnet which could be
explained in the framework of a simple ``centrifugal model''.
There is one exception only, $m=-1$, where the local mode exists
for every soliton radius.

Our investigations can be applied to the description of the
eigenmodes of magnetic dots. In the present paper we develop the
theory of local modes on the soliton background; it could be a
good guide for the study of the normal modes in the vortex--state
magnetic dots. Our theory is constructed for the soliton in the
model (\ref{eq:L}), where the soliton is stabilized by the
internal precession. It is clear that such a model cannot
guarantee the quantitative correspondence with the case of
vortex--state magnetic dots, where the static soliton structure
is stabilized by the magnetic--dipole interaction. We did not
consider this type of interaction in the paper, as it is
difficult to account for. Nevertheless we believe that the main
features of the problem studied above are generic. First of all,
we expect the appearance of modes with anomalously small
frequencies, e.g. the mode of the translational oscillations of
the vortex centre. The non--zero frequency of this mode is caused
by the interaction with the boundary only. Secondly, doublets with
$m=\pm n$, $n>1$, should appear, when the radius of the vortex is
rather large.

We have used the results on the local modes for the most
interesting case of the translational modes with $|m|=1$ to
describe the soliton motion in the infinite--size magnet: it is
possible to identify the soliton mass, which is finite due to the
localized soliton structure. In contrast to both the
Belavin--Polyakov soliton in the isotropic FM and the magnetic
vortex in the easy--plane FM, the soliton motion in our case of
the easy--axis FM is similar to the motion of a finite mass
charged particle in a magnetic field.

Thus, to the best of our knowledge, the localized precessional
soliton in the easy--axis FM is the first example of a 2d
topological magnetic soliton with truly particle--like properties.

\acknowledgments

D.D.Sh thanks the University of Bayreuth, where part of this work
was performed as a post--doc of the Graduiertenkolleg
``Nichtlineare Spektroskopie und Dynamik'', for kind hospitality.
Work at Kiev was partially supported by INTAS--97-31311 Grant.

% ###################################################################

%\bibliographystyle{prsty}
%\bibliography{soliton}

\begin{thebibliography}{99}
\bibitem[*]{DDSh_address}
Permanent Address: National Taras Shevchenko University of Kiev,
03127 Kiev, Ukraine. Electronic address:
Denis\_Sheka@mail.univ.kiev.ua

\bibitem{Berezinsky72}
V.~L. Berezinski\u\i, Sov. Phys JETP {\bf 34},  610  (1972).

\bibitem{Kosterlitz73}
J.~M. Kosterlitz and D.~J. Thouless, J. Phys. {\bf C 6},  1181
(1973).

\bibitem{Belavin75}
A.~A. Belavin and A.~M. Polyakov, JETP Lett. {\bf 22},  245
(1975).

\bibitem{Currie80}
J.~F. Currie, J.~A. Krumhansl, A.~R. Bishop, and S.~E.
Trullinger, Phys. Rev.
  {\bf B 22},  477  (1980).

\bibitem{Mikeska91}
H.~J. Mikeska and M. Steiner, Adv. Phys. {\bf 40},  191  (1991).

\bibitem{Wiesler94}
D.~D. Wiesler, H. Zabel, and S.~M. Shapiro, Z. Phys. {\bf B 93},
277  (1994).

\bibitem{Ivanov95e}
B.~A. Ivanov and A.~K. Kolezhuk, Low Temp. Phys. {\bf 21 (4)},
275  (1995).

\bibitem{Boucher87}
J.~P. Boucher, G. Rius, and Y. Henry, Europhys. Lett. {\bf 4},
1073  (1987).

\bibitem{Mertens87}
F.~G. Mertens, A.~R. Bishop, G.~M. Wysin, and C. Kawabata, Phys.
Rev. Lett.
  {\bf 59},  117  (1987).

\bibitem{Mertens89}
F.~G. Mertens, A.~R. Bishop, G.~M. Wysin, and C. Kawabata, Phys.
Rev. {\bf B 39},  591  (1989).

\bibitem{Kosevich90}
A.~M. Kosevich, B.~A. Ivanov, and A.~S. Kovalev, Phys. Rep. {\bf
194},  117  (1990).

\bibitem{Waldner83}
F. Waldner, J.~Magn. Magn. Mater. {\bf 31---34},  1203  (1983).

\bibitem{Waldner86}
F. Waldner, J.~Magn. Magn. Mater. {\bf 54---57},  873  (1986).

\bibitem{Waldner92}
F. Waldner, J.~Magn. Magn. Mater. {\bf 104---107},  793  (1992).

\bibitem{Zaspel93}
C.~E. Zaspel, Phys. Rev. {\bf B 48},  926  (1993).

\bibitem{Zaspel95}
C.~E. Zaspel, T.~E. Grigereit, and J.~E. Drumheller, Phys. Rev.
Lett. {\bf 74},
   4539  (1995).

\bibitem{Zaspel96}
C.~E. Zaspel and J.~E. Drumheller, International Journal of
Modern Physics {\bf
  10},  3649  (1996).

\bibitem{Subbaraman98}
K. Subbaraman, C.~E. Zaspel, and J.~E. Drumheller, Phys. Rev.
Lett. {\bf 80}, 2201  (1998).

\bibitem{Ivanov99}
B.~A. Ivanov, V.~M. Muravyov, and D.~D. Sheka, JETP {\bf 89},
583  (1999).

\bibitem{Walliser99}
H.~Walliser, G.~Holzwarth, hep-ph/9907492 (unpublished).

\bibitem{Ivanov98}
B.~A. Ivanov, H.~J. Schnitzer, F.~G. Mertens, and G.~M. Wysin,
Phys. Rev. {\bf B 58},  8464  (1998).

\bibitem{Wysin94}
G.~M. Wysin, Phys. Rev. {\bf B 49},  8780  (1994).

\bibitem{Wysin96}
G.~M. Wysin and A.~R. V{\"o}lkel, Phys. Rev. {\bf B 54},  12921
(1996).

\bibitem{Ivanov95g}
B.~A. Ivanov, JETP Lett. {\bf 61},  917  (1995).

\bibitem{Ivanov98a}
B.~A. Ivanov and V.~M. Muravyov, Low Temp. Phys. {\bf 24},  510
(1998).

\bibitem{Abdullaev99}
F.~K. Abdullaev, R.~M. Galimzyanov, and A.~S. Kirakosyan, Phys.
Rev. {\bf B
  60},  6552  (1999).

\bibitem{Ivanov96}
B.~A. Ivanov, A.~K. Kolezhuk, and G.~M. Wysin, Phys. Rev. Lett.
{\bf 76},  511
  (1996).

\bibitem{Hillebrands97}
B.~Hillebrands, C.~Mathieu, C.~Hartmann, M.~Bauer, O.~Buettner,
S.~Riedling, B.~Roos, S.~O.~Demokritov, B.~Bartenlian,
C.~Chappert, D.~Decanini, F.~Rousseaux, E.~Cambrill,
A.~M\"{u}ller, B.~Hoffmann, and U.~Hartmann, J.~Magn. Magn.
Mater. {\bf 175}, 10 (1986).

\bibitem{Grimsditch98}
M.~Grimsditch, Y.~Jaccard, and I.~K.~Schuller, Phys. Rev. {\bf B
58},  11539 (1998).

\bibitem{Usov93}
N.~A.~Usov and S.~E.~Peschany, J.~Magn. Magn. Mater. {\bf 118},
L290 (1993).

\bibitem{Mathieu98}
C.~Mathieu, J.~Jorzick, A.~Frank, S.~O.~Demokritov, A.~N.~Slavin,
B.~Hillebrands, B.~Bartenlian, C.~Chappert, D.~Decanini,
F.~Rousseaux, and E.~Cambrill, Phys. Rev. Lett. {\bf 81}, 3968
(1998).

\bibitem{Guslienko00}
K.~Yu.~Guslienko and A.~N.~Slavin, J.~Appl. Phys., {\bf 87}, 6337
(2000).



\bibitem{Bar'yakhtar93}
V.~G. Bar'yakhtar and B.~A. Ivanov, Sov. Sci. Rev. Sec.A. ---
Phys. Reviews ed.
  by I. Khalatnikov, Amsterdam {\bf 16},  3  (1993).

\bibitem{Voronov83}
V.~P. Voronov, B.~A. Ivanov, and A.~K. Kosevich, Sov. Phys JETP
{\bf 84},  2235
   (1983).

\bibitem{Ivanov86}
B.~A. Ivanov and V.~A. Stephanovich, Sov. Phys JETP {\bf 64},
376  (1986).

\bibitem{Ivanov99b}
B.~A. Ivanov and A.~A. Zhmudsskii, JETP {\bf 88},  833  (1999).

\bibitem{Harrison66}
W.~A.~Harrison, {\em Pseudopotentials in the Theory of Metals}
(New York, 1966).

\bibitem{Wysin96a}
G.~M. Wysin, Phys. Rev. {\bf B 54},  15156  (1996).

\bibitem{Mertens97}
F.~G. Mertens, H.~J. Schnitzer, and A.~R. Bishop, Phys. Rev. {\bf
B 56},  2510 (1997).

\bibitem{Mertens99}
F.~G. Mertens and A.~R. Bishop,  in {\em Nonlinear Science at the
Dawn of the
  21th Century}, edited by P.~L. Christiansen and M.~P. Soerensen
  (Springer--Verlag, Berlin, 1999).

\bibitem{Ivanov89}
B.~A. Ivanov and V.~A. Stephanovich, Phys. Lett. {\bf A 141}, 89
(1989).


\end{thebibliography}

% ###################################################################

% Tables and figure captions:
\clearpage

\begin{table}
\caption{Fitting of the trial solution (\ref{eq:theta-trial}) to
the exact numerical solution for $\theta_0(r)$. Radius of the
soliton is defined from the condition $\theta_0(R)=\pi/2$.} %
\label{table:fitting} %
\begin{tabular}{cccc}
Frequency of                  & Soliton         & Fitting para- & Std. error, \\ %
precession, $\Omega/\omega_0$ & radius, $R/\Delta_0$   & meter,
$R^{\text{(fit)}}/\Delta_0$   & $\times 10^{-2}$ \\
\hline
$0.05$  & $20.38$ & $20.38$ & $0.6$ \\ %
$0.10$  & $9.95$  & $9.94$  & $1.2$ \\ %
$0.15$  & $6.58$  & $6.58$  & $1.9$ \\ %
$0.20$  & $4.88$  & $4.87$  & $2.6$ \\ %
$0.25$  & $3.82$  & $3.81$  & $3.2$ \\ %
$0.30$  & $3.09$  & $3.07$  & $3.6$ \\ %
$0.40$  & $2.07$  & $2.07$  & $4.0$ \\ %
$0.50$  & $1.35$  & $1.37$  & $3.8$ \\ %
$0.60$  & $0.81$  & $0.85$  & $3.2$ \\ %
$0.70$  & $0.42$  & $0.44$  & $2.2$ \\ %
$0.80$  & $0.15$  & $0.16$  & $1.1$ \\ %
$0.85$  & $0.07$  & $0.07$  & $0.9$ \\ %
\end{tabular}
\end{table}

\begin{table}
\caption{Critical parameters for some lower modes with $|m|>1$.
Remember that local modes always exist for $m=0,\pm1$.} %
\label{table:Rc} %
\begin{tabular}{cccc}
$m$ & $R_m^{\text{c}}/\Delta_0$, & $\Omega_m^{\text{c}}/\omega_0$,
& $\Omega_m^{\text{c}}/\omega_0$, \\
& data & data & from Eq.~(\ref{eq:Omega^c-est})\\ \hline %
%$0,\pm1$ & absent & absent & absent  \\ %
$-2$     & $1.52$ & $0.47$ & $0.45$  \\ %
$+2$     & $2.24$ & $0.38$ & $0.38$  \\ %
$-3$     & $2.73$ & $0.33$ & $0.30$  \\ %
$+3$     & $3.22$ & $0.29$ & $0.28$  \\ %
\end{tabular}
\end{table}

\end{document}